\documentclass[twocolumn]{aastex63}

\usepackage{amsmath}
\usepackage{graphicx}
\usepackage{wrapfig}
\usepackage[figuresleft]{rotating}
\usepackage{multirow}
\usepackage{float}

\submitjournal{ApJ (version closer to published in ApJ)}

\shorttitle{N-body Fornax dSph galaxy modeling}
\shortauthors{G. Shchelkanova et al.}

\begin{document}

\title{N-body self-consistent stars-halo modeling of the Fornax dwarf galaxy}

\correspondingauthor{Galina Shchelkanova}
\email{mikgalina@gmail.com}

\author{Galina Shchelkanova}
\affiliation{NRC ``Kurchatov Institute'' - ITEP, 117218 Moscow, Russia}

\author{Kohei Hayashi}
\affiliation{Astronomical Institute, Tohoku University, Aoba-ku, Sendai 980-8578, Japan}
\affiliation{Institute for Cosmic Ray Research (ICRR), The University of Tokyo, Tokyo 277-8582, Japan}

\author{Sergei Blinnikov}
\affiliation{NRC ``Kurchatov Institute'' - ITEP, 117218 Moscow, Russia}
\affiliation{Kavli Institute for the Physics and Mathematics of the Universe (Kavli IPMU, WPI),\\
 The University of Tokyo, Chiba 277-8583, Japan}
\affiliation{Sternberg Astronomical Institute, Lomonosov Moscow State University, Universitetsky prospekt 13, 119234, Moscow, Russia}



\begin{abstract}

We present nearly self-consistent stellar-halo models of the
Fornax dwarf spheroidal galaxy associated with the Milky Way galaxy.
Such galaxies are
dominated by dark matter and have almost no gas in the system.
Therefore,
they are excellent objects for N-body modeling
that takes into account visible and dark matter halo components.
In order to model the dark matter halo inferred from
the analysis of the measured velocities of Fornax's stars, 
we constructed several self-consistent quasi-equilibrium 
models based on two source code sets.
One of them (GalactICS Software, NEMO) deals with the
self-consistent distribution function modeling 
which depends
on energy $E$ and 
vertical component of the
angular momentum $L_z$.
The other is included in the AGAMA framework and is
 based on Schwarzschild's calculation of orbits.
It can reproduce the non-spherical 
self-consistent structure of Fornax
as the weighted sum of orbit contributions
to the galactic density
even though the inferred dark halo parameters come 
from Jeans analysis which does not require that 
any distribution functions should be positive.
To guess the parameters which make the N-body
models close to the visible object we use the 
stellar-dark matter model of the Fornax galaxy
based on hydrodynamic axisymmetric Jeans equations taking into 
account the velocity anisotropy parameter.
Then we 
studied the evolution of the models by performing
N-body simulations with the falcON code
in order to test their stability.
The variability of the model parameters
over time was obtained during simulations.
The AGAMA models show the best agreement of the resulting 
velocity dispersion profiles with the observed data.

\end{abstract}


\section{Introduction} \label{sec:intro}
Dwarf spheroidal (dSph) galaxies associated with the Milky Way
(MW) and M31 galaxies are the best probes
for studying the properties of dark matter. This is because 
these galaxies are largely dark matter-dominated
objects with dynamical mass-to-light ratios of 10 to 
1000~\citet{Mateo1998},~\citet{Gilmore2007},~\citet{LDwf}.
Moreover, in the context of bottom-up formation scenarios based
on $\Lambda$-cold dark matter ($\Lambda$CDM) theory,
these galaxies are the building blocks of more massive galaxies,
and thus studying their properties
and evolution is important to understanding galaxy 
formation ~\citep[e.g.,][]{Tolstoy2009}.

The hypothesis on the existence of dark matter (DM) was put 
forward by~\citet{Zwicky1933}
to explain the virial paradox in the Coma cluster.
Later~\citet{Babcock1939} found the growth of the rotation 
curve of M31 in its outer parts  and concluded 
that a large amount of
invisible mass was present in it.
The hypothesis about DM was revived by~\citet{Einasto1974}
and~\citet{Ostriker1974} in their studies 
on the rotation curves of galaxies, which appeared to be 
mostly flat.
The flat rotation curves could be artificially 
explained without DM
by the surface density in the disk
being inversely proportional to the distance from the center, 
as established  by~\citet{Mestel1963}.
Nevertheless, the strongest arguments in favor of 
the existence of DM come from the need to explain the 
Large-scale structure formation given the small amplitudes 
of perturbations in the cosmic microwave background 
radiation (CMB).


$\Lambda$CDM theory is a concordance model in modern cosmology 
that gives an excellent description of the 
CMB~\citep[e.g.,][]{Planck2014},
large-scale structure~\citep[e.g.,][]{Tegmark2004}, 
and 
the accelerating expansion of the 
Universe~\citep[e.g.,][]{accelerationSNIa}.
On the other hand, the observational studies on the galactic and 
sub-galactic scales have turned up several
controversial issues that continue to challenge the 
$\Lambda$CDM paradigm.
The core-cusp problem is one of the open questions in 
the $\Lambda$CDM picture: the cosmological 
$\Lambda$CDM-based pure dark matter
simulations predict that the dark halos on all mass scales 
have cusped dark matter density profiles.
On the other hand, studies of the HI gas rotation curves of 
low surface brightness galaxies and the stellar
kinematics of dwarf galaxies testify in favour of
shallower or cored density profiles of galaxies' dark halos~\citep[e.g.,][]{deBlok2001,Gilmore2007}.

Another way to explain the lack of visible matter was proposed 
by~\citet{Milgrom1983} who
hypothesized that there was a characteristic acceleration
below which Newton's law of gravitation was invalid. 
This is one of the varieties
of Modified Gravity (MG) that was subsequently developed, 
for example, by~\citet{Bekenstein2004}
and~\citet{Bekenstein2010} as the tensor-vector scalar theory, 
or TeVeS. A variety of MG models has been developed 
until now~\citep[see][]{MG2017}, but the $\Lambda$CDM 
theory is still the leading paradigm of modern cosmology 
because the MG is not able to explain 
the alternative formation paths of the large-scale
structure of the Universe as well as
the entire set of observations for galaxies and their clusters.
For a review, see for example,~\citet{Read2019}, where 
different star formation history causes different central 
DM densities.

Early dark matter study of the Milky Way dSphs
was done by~\citet{Lin}. More recent research can be 
classified as based on Jeans equations, 
distribution functions (DFs),
action-based DFs, and Schwarzschild's orbit-based
methods~\citep[for a review see][]{Battaglia13}.
In this work, we focus on Fornax dSph, which is well-studied 
for its dark matter halo structure.
Moreover, this galaxy is suggested to have a core-like 
dark matter density profile,
which behaves like $\rho_{\mathrm{DM}} \sim r^{\alpha}$, where 
$\alpha=-0.22$ estimated by~\citet{H16}.
Thus, the galaxy can assess the core-cusp problem in the 
$\Lambda$CDM models.

The non self-consistent (stars in the field of DM)
spherical Jeans equation approaches
were employed taking into account 
velocity anisotropy for Fornax
by~\citet{Gilmore2007,Penarrubia2008,Strigari,W09ApJ,Salucci2012,Read2019}.
Axisymmetric models were constructed by~\citet{HC12,H16}.
The DF-based modeling for the stellar component
in the field of a parametrized spherical DM potential
for Fornax dSph galaxy was done
by~\citet{Wu2007} and~\citet{Amorisco}.
But all these models are not self-consistent
in contrast to the method of~\citet{KD95} 
who iteratively solved the Poisson equation taking into 
account the stellar and DM DFs in the common gravitational 
potential. Their method is realized in the framework
of the GalactICS Software.

Several methods based on
the action-based distribution functions
have been 
developed~\citep[for a review see][models for globular clusters]{Sanders16,Jeffreson17}.
The implementation for the Fornax galaxy was done by~\citet{Pascale},
but their models were spherically symmetric.

Schwarzschild's modeling~\citep{Schwarzschild,Richstone}
was applied to Fornax, Sculptor, Carina and Sextans 
dSph galaxies by~\citet{BreddelsHelmi13}.
The non-spherical light distribution in the spherical DM field
was taken into account by~\citet{JG12}.
The spherically symmetric Schwarzschild model of the Fornax dSph
galaxy was recently done by~\citet{Kowalczyk19}.
In this work, we perform two kinds of self-consistent 
stellar-halo modelings of the Fornax dSph,
which are the DF-based one
constructed by~\citet{KD95}
and the orbit-based one coded within the AGAMA framework~\citep{Vasiliev2018}
without assumptions on spherical symmetry.
To this end, we utilize the results of the dark matter structures
in the Fornax galaxy analyzed by~\citet{HC12}
and~\citet{H16}.
In order to trace the N-body dynamical evolution of our 
models and to check their self-consistency and stability,
we used the code by~\citet{Den} named falcON.

In this work, we shall find the set of parameters like 
visible and DM masses, characteristic radii, the radial 
behaviour of the DM profile near the center and at the periphery.
Our goal is to make the modeled
dynamical characteristics satisfy the observed ones, 
such as the velocity dispersion profile.
We will perform our investigation in the framework of the 
standard CDM model.
The construction of a feasible evolutionary stable galaxy model 
is important to find constrains on possible DM candidates 
as is shown by~\citet{Gonzalez-Morales} and \citet{Safarzadeh}.

\section{Stellar-halo model}

N-body simulations have begun long ago, see~\citet{1983MNRAS.204..891K}.
We construct equilibrium systems (coordinates, masses and 
velocities of particles)
by two methods and then we use the falcON code
in order to follow the N-body evolution of the systems.

While constructing our systems, we adjust their stellar
surface density profiles the observations
such as King~\citep{K62} or Plummer~\citep{Plum} profiles.
In order to compare our kinematic characteristics to the data,
we use the observational velocity dispersion
projected onto the line-of-sight profiles done by~\citet{HC12}
from the data of~\citet{W09AJ}.

Our nearly self-consistent stars-halo models 
of the Fornax galaxy are constructed using
the DF-based method of~\citet{KD95}
and the orbit-based Schwarzschild's method by~\citet{Vasiliev2018}.
This is a step forward after previous studies
of this object, which were based on the more approximate 
Jeans equations
or spherically symmetric DF or action-based approach.
The fitting of the velocity dispersion profile to the data
was done only for the mock galaxy by~\citet{Vasiliev2020}
and has not been done for a real astrophysical object.
We did not perform the fitting procedure but 
instead constructed
many axisymmetric models with different assumptions
about stellar and DM density distributions that satisfy the 
velocity data rather well.
From the set of these  models, we pick up one model developed 
by the GalaxtICS Software NEMO code and the other
two  models by the Schwarzchild’s based code in the AGAMA framework.
We report about these models in order to:
\begin{itemize}
\item show the best NEMO and AGAMA models 
(number 1 and 2 in Table~\ref{Models_params});
\item compare two different approaches and codes
(NEMO-model 1 is analogous to AGAMA-model 3);
\item study the process of relaxation to equilibrium 
for different values (NEMO-models are changed
more during this process than the AGAMA ones);
\item trace the magnitude of change of the previous 
Jeans velocity dispersion fitting in the field of
the axisymmetric prolate DM halo 
found by~\citet{H16} (AGAMA model~3).
\end{itemize}

\section{Density models and parameters} \label{section_Data}

For the density profile of the stellar component of the galaxy,
we use spherically symmetric King
(AGAMA model~2  in Table~\ref{Models_params}) 
and axially symmetric Plummer
(NEMO model~1 and AGAMA model~3) profiles.

For the King profile, we employ formulas:
\begin{multline}
\label{King_dist}
\rho_{\rm King}(r)=\begin{cases}
\rho_b \left( \frac{1}{\sqrt{1+(r/r_c)^2}} - \frac{1}{\sqrt{1+(r_t/r_c)^2}} \right)^2,&\text{$r\leq r_t$;}\\
0,&\text{$r > r_t$,}
\end{cases}
\end{multline}
where $r$ is the spherical radial coordinate,
$\rho_b,r_c$ and $r_t$ are the central stellar density,
the core and tidal radius, respectively.
In order to calculate the density parameter $\rho_b$
we use formulas~(\ref{rhob_inTermsof_Sigma}) 
in Appendix~\ref{King_surface}.
We take into account the
central surface brightness $\Sigma_0=11.0 \; L_\odot/\mbox{pc}^2$ 
and the core and tidal radii as $r_c=0.53 \; \mbox{kpc}, r_t=1.66 \; \mbox{kpc}$.
The value of $\Sigma_0$ is slightly shifted inside the error bar
with respect to the data of~\citet{IH}, and the values of the
radii are not inside their error bars (see table~\ref{Data_visible_params}),
but the best fit overall has been obtained with these values.
As we shall see further from Figure~\ref{start_rhoLum_model1}
this stellar density profile does not differ a lot from the profile of~\citet{IH}.

\begin{deluxetable}{cccccccc}
\tablecaption{Observed parameters of the Fornax galaxy.\label{Data_visible_params} }
\tablehead{
\colhead{$\Sigma_0$}                         &\colhead{$D$}                      &\colhead{$r_c$}           &\colhead{$r_t$}           &\colhead{$r_{\rm half}$ }&\colhead{$M$}             \\
\colhead{$\left[ L_\odot/\mbox{pc}^2\right]$}&\colhead{$\left[\mbox{kpc}\right]$}&\colhead{$\left['\right]$}&\colhead{$\left['\right]$}&\colhead{$[\mbox{kpc}]$ }&\colhead{$[10^6 M_\odot]$}
}
\startdata
                   $15.7 \pm 5.1 $           &   $ 147 $                         &      $ 17.6 \pm 0.2$     &      $ 69.1 \pm 0.4$     &           $ 0.668 $     &         $ 20$             \\
\enddata
\tablecomments{The central surface brightness ($\Sigma_0$) from~\citet{IH},
the distance to the Fornax galaxy ($D$)
from~\citet{Piet},
the structural parameters ($r_c$ and $r_t$) for the King formulas~\ref{King_dist}
from~\citet{Bat}, the
half-light radius ($r_{\rm half}$) and luminous mass ($M$) of the Fornax galaxy
from~\citet{LDwf}.
}
\end{deluxetable}

\begin{deluxetable}{ccDcccl}
\tablecaption{Modeling parameters of the Fornax galaxy.\label{Models_params} }
\tablehead{
\colhead{Model}&
\colhead{\multirow{2}*{\rotatebox{90}{code}}}&
\multicolumn2c{$\alpha$}&
\colhead{\multirow{2}*{\rotatebox{90}{stellar}\rotatebox{90}{profile}}}
                                     & \colhead{$M_p$}             &\colhead{$b_p$}           &\colhead{$q$}           \\
& &\multicolumn2c{}& &\colhead{$\left[10^6 M_\odot\right]$}        &\colhead{$[\mbox{kpc}]$}  &\colhead{}
}
\decimals
\startdata
1      & N & -0.22 &\multirow{2}*{\rotatebox{30}{Plummer}}%
                     &               $20.0$ &       $0.668$        &          -         \\ 
3      & A & -0.22 & &               $20.0$ &       $0.668$        &         $0.66$       \\ 
\hline
& &\multicolumn2c{}&           &          $\Sigma_0$               &         $r_c$          &    $r_t$               \\
& &\multicolumn2c{}&           &$\left[ L_\odot/\mbox{pc}^2\right]$&   $\left['\right]$     &    $\left['\right]$    \\
\hline
2      & A &  0.0  & King      &                $11.0$             &            $12.4$      &             $38.8$   \\
\enddata
\tablecomments{Different visible parameters, DM profile exponent $\alpha$,
and the type of the {\bf modeling} code:
``N''~-- NEMO DF-based by~\citet{KD95};
``A''~-- orbit-based AGAMA by~\citet{Vasiliev2018}.
For the model~{\bf 1} we have gotten the Plummer-like density profile
for the equatorial plane,
it is the function of the combined
gravitational potential, which is axially symmetric, but
the equipotentials are not ellipsoids.
}
\end{deluxetable}

For the oblate Plummer profile, we use the following
function $\rho_{\rm Plummer}(R,z)$ of the cylindrical coordinates as in~\citet{H16}:
\begin{multline}
\label{Plum_dist}
\begin{cases}
&\rho_{\rm Plummer}(R,z)=\rho_p(m_\star)= \frac{3 M_p}{4 \pi {b_p}^3} \left[ 1 + \frac{{m_\star}^2}{{b_p}^2} \right]^{-5/2} \text{, } \\
&{m_\star}^2 = R^2 + \frac{z^2}{q^2}\text{ ,}
\end{cases}
\end{multline}
with half-light radius $b_p$
calculated by~\citet{Werr} as the half surface
brightness of the King profile. The same value for the Plummer profile
 was used by~\citet{H16}.
As for the mass parameter for the Plummer profile $M_p$, we use the mass
of the Fornax galaxy from~\citet{LDwf}. These values
are listed in table~\ref{Data_visible_params}.
The oblateness parameter $q$ is calculated from the apparent 
axial ratio $q'$ and the galaxy inclination $i$ is
taken from \citet{H16} (see table~\ref{Data_Kohey_params}) by the formula:
\begin{eqnarray}
\label{q_ap}
q=\sqrt{{q'}^2-\cos^2 i}/\sin i \; ,
\end{eqnarray}
with the parameters listed in Table~\ref{Data_Kohey_params}.
The density profile of the DM halo was also taken from \citet{H16}.
It is a function $\rho(R,z)$ of the cylindrical coordinates with parameters 
also listed in Table~\ref{Data_Kohey_params}:
\begin{multline}
\label{rhoDM_distribution}
\rho(R,z)=\rho(m) =  \\
\rho_0 \left( \frac{m}{b_{\rm halo}} \right)^\alpha
\left[ 1 + \left( \frac{m}{b_{\rm halo}} \right)^2 \right]^{-(\alpha + 3)/2}, \;
m^2 = R^2 + \frac{z^2}{Q^2} \; ,
\end{multline}
where $b_{\mathrm{halo}}, \alpha$ and $Q$
are the scale length, the inner slope, and the axial ratio
of dark matter density profile, respectively.
We also tried the cored DM profile as was done in~\citet{HC15}
but for the prolate form of the halo (AGAMA model~2).

\begin{splitdeluxetable}{cRRRBRRr}
\tablecaption{Hydrodynamical parameters.\label{Data_Kohey_params}}
\tablehead{
\colhead{}&\colhead{$\alpha$}&\colhead{$\log_{10}\left(\rho_0\right)$}    &\colhead{$\log_{10}\left(b_{\rm halo}\right)$}&\colhead{$Q$}&\colhead{$i$}    &$q'$ \\
\colhead{}&\colhead{}        &\colhead{$\left[M_\odot/\mbox{pc}^3\right]$}&\colhead{$[\mbox{pc}]$}                       &\colhead{}   &\colhead{$[deg]$}&\colhead{}
}
\startdata
(1)&-0.22^{+0.14}_{-0.22}& -1.06 \pm 0.15 &  2.79^{+0.16}_{-0.15} & 1.11^{+0.66}_{-0.53}  &71.85^{+11.56}_{-15.34}       &    $0.7$ \\
\enddata
\tablecomments{Values from~\citet{H16}~-- (1).}
\end{splitdeluxetable}

Taking into account all this data and varying parameters
we constructed three models. The diversity of parameters for
the stellar component and two variants of the exponent
of the DM profile $\alpha$ are listed in Table~\ref{Models_params}.
The model numbered 1 is implemented by
the DF-based mkkd95 code (GalactICS Software, NEMO)
designed by~\citet{KD95} and the last two ones by the orbit-based
Schwarzschild's method implemented in the AGAMA
framework designed for constructing equilibrium models by~\citep{Vasiliev2018}.
For all these methods we used $G=1$ as the gravitational constant,
$10^6 M_{\odot}$ as the mass unit, and $ 1\mbox{ kpc}$ as the unit of distance.
Then for the time unit we have:
\begin{eqnarray}
\label{time_unit}
[T] = \frac{\mbox{kpc}^\frac{3}{2}}{\left( G \times 10^6 M_\odot \right)^\frac{1}{2}}
    = 0.47 \; \mbox{Gyr} \; .
\end{eqnarray}

Model~1 relies on the
shallow DM profile with $\alpha=-0.22$ and
uses a Plummer-like profile for the visible component of the galaxy.
The density distribution depends on the combined gravitational 
potential, so it is axially symmetric and coincides with the Plummer 
analytical density profile at the equatorial galaxy plane ($z=0$),
but the isopycnic surfaces are not ellipsoids.
The model does not take into account kinematic constraints
for the visible part of the galaxy obtained by~\citet{H16}.
The parameter $\beta_z$ stands for such a constraint.
This is a velocity anisotropy parameter:
\begin{equation}
\label{beta_z}
\beta_z = 1 - \overline{{v_z}^2}/\overline{{v_R}^2} \; .
\end{equation}
For the Fornax galaxy $-\log_{10}(1-\beta_z) = 0.28^{+0.10}_{-0.11}$.

The AGAMA orbit-based
code has three different parameters
to constrain the velocity anisotropy in the solution:
$\beta$, $\beta_z$, and $\kappa$.
The $\beta$ parameter is the spherical anisotropy index:
\begin{equation}
\beta = 1 - {\sigma_\tau}^2/(2*{\sigma_r}^2) \; ,
\end{equation}
where $\sigma_\tau$ is the tangential velocity dispersion,
${\sigma_\tau}^2 = {\sigma_\phi}^2 + {\sigma_\theta}^2$,
and $\sigma_r$ is dispersion of velocity along the spherical radius.
And setting
\begin{equation}
\kappa = 1.0 \; \mbox{i.e.} \; \sigma_\phi=\sigma_R \; ,
\end{equation}
where $\sigma_R$ is the dispersion of velocity along the cylindrical radius.

Model~2 assumes the visible King profile and the cored DM profile.
Its King profile differs slightly from the observed one, but the model has
the best reproduction of the velocity dispersion profile
(see the Results section).
The velocity anisotropy is expressed by the $\beta$ parameter.
Model~3 describes the cusped DM profile and the visible Plummer 
profile with $\beta_z$ and $\kappa$ parameters for the kinematic constrains.

For all our N-body models we used $10^6$ stellar
particles and
$1.5 \times 10^6$ DM halo points.
For the AGAMA Schwarzschild's orbit-based modeling we used
$25000$ orbits for each component.
For the falcON runs we used all the default parameters except
$k_{max}=6$ ($\tau_{max}=(1/2)^{k_{max}}$) and softening length
$\epsilon=0.1$ (for the code notation see Appendix~\ref{code_notations}).

\subsection{DF-based NEMO modeling} \label{Nemo_modeling}

For the first sample of our nearly self-consistent stellar-halo modeling
(Table~\ref{Models_params}),
we use the bulge and DM components of the NEMO code
developed by 
\citet{KD95}.
For the visible component in this code the 
bulde with King's density profile is used
that has the DF described in~\citep{KD95} by the equation:
\begin{multline}
\label{KD95_eq1}
f_{\rm bulge}(E) = \\
\begin{cases}
\rho_b (2\pi{\sigma_b}^2)^{-\frac{3}{2}}e^{ \frac{\Psi_0-\Psi_c}{{\sigma_b}^2}} \left( e^{-\frac{E-\Psi_c}{{\sigma_b}^2}} - 1 \right) &\text{$E<\Psi_c$,}\\
0 &\text{$E\geq\Psi_c$,}
\end{cases}
\end{multline}
and the density distribution in a potential $\Psi$ is described by 
the equation: 
\begin{multline}
\label{KD95_eq2}
\rho_{\rm bulge}(\Psi) =
    \rho_b \left[
           e^{\frac{\Psi_0-\Psi}{{\sigma_b}^2}} \mbox{erf}\left(\frac{\sqrt{\Psi_c-\Psi}}{\sigma_b} \right)
           \right.\\
           \left.
         - \frac{1}{\sqrt{\pi}} e^{\frac{\Psi_0-\Psi}{{\sigma_b}^2}}
             \left(
                   2 \frac{\sqrt{\Psi_c-\Psi}}{\sigma_b}
                   + \frac{4}{3}\frac{\left(\Psi_c-\Psi \right)^\frac{3}{2}}{{\sigma_b}^3}
             \right)
           \right]
\end{multline}
This bulge density distribution follows the equipotential surfaces, so it is neither spherical nor
ellipsoidal with the given oblateness. It has three parameters: $\Psi_c$, $\rho_b$ and $\sigma_b$.

The DM component construction is based on
the lowered Evans distribution~\citet{Evans93}
also described in~\citet{KD95}. The DF for DM is presented as
\begin{multline}
\label{KD95_eq3}
f_{\rm halo} (E,{L_z}^2) = \\
\begin{cases}
\left[ \left( A {L_z}^2 + B \right) e^{-\frac{E}{{\sigma_0}^2}} + C \right] \left[ e^{-\frac{E}{{\sigma_0}^2}} - 1 \right] &\text{$E<0$,}\\
0 &\text{$E\geq 0$,}
\end{cases}
\end{multline}
and the density profile is given by 
\begin{multline}
\label{KD95_eq4}
\rho_{\rm halo} (R,\Psi) = \frac{1}{2} {\pi}^\frac{3}{2} {\sigma_0}^3
        \left( A R^2 {\sigma_0}^2 + 2 B \right)
        \mbox{erf} \frac{\sqrt{-2\Psi}}{\sigma_0}
        e^{\frac{-2\Psi}{\sigma_0} } \\
   + (2\pi)^\frac{3}{2} {\sigma_0}^3
     \left(C-B-AR^2{\sigma_0}^2\right)
     \mbox{erf} \frac{\sqrt{-\Psi}}{\sigma_0}
     e^\frac{-\Psi}{{\sigma_0}^2} \\
+ \pi \sqrt{-2\Psi} \left[
                          {\sigma_0}^2
                              \left( 3A{\sigma_0}^2R^2 + 2B - 4C \right)
                     \right.\\
                     \left.
                         +\frac{4}{3} \Psi \left( 2C - A{\sigma_0}^2R^2\right)
                    \right]\; .
\end{multline}
Parameters $A$, $B$, $C$ are expressed by the velocity and density scales
$\sigma_0$ and $\rho_1$, the halo core radius $R_c$, and the 
flattening parameter $q$ as follows (see~\citet{KD94}):
\begin{eqnarray}
\label{A}
&A = \frac{8(1-q^2)G{\rho_1}^2}{\sqrt{\pi}q^2{\sigma_0}^7}\\\label{B}
&B = \frac{4 {R_c}^2 G {\rho_1}^2}{\sqrt{\pi}q^2{\sigma_0}^5}\\\label{C}
&C = \frac{(2q^2-1)\rho_1}{(2\pi)^\frac{3}{2}q^2{\sigma_0}^3}
\end{eqnarray}
The density scale $\rho_1$ is replaced in~\citet{KD95} by $R_a$:
\begin{eqnarray}
\label{r_a}
R_a = \left( \frac{3}{2\pi G\rho_1} \right)^\frac{1}{2} \sigma_0 e^{\frac{\Psi_0}{2{\sigma_0}^2}}
\end{eqnarray}

For the DM NEMO model we need 5 parameters:
\begin{itemize}
\item $q$~--axial ratio, an optional flattening parameter for the potential ;
\item $\Psi_0$~-- central potential;
\item $v_0 = \sqrt{2}\times\sigma_0$, where $\sigma_0$ is  the central  velocity  dispersion;
\item $R_a$~-- the radius at which the halo rotation curve, if continued at its $r=0$ slope, would reach the value $\sqrt{2}\sigma_0$,
a scaling radius for the halo;
\item ${r_{ck}}^2 = \dfrac{{R_c}^2}{{R_K}^2}$~-- a core smoothing parameter~--
the ratio of the core radius ($R_c$) to the derived King radius ($R_K$).
This is the radius
 at which the gravitational potential has risen
by about $2{\sigma_0}^2$ over its central value, provided that the 
potential well depth is above $2{\sigma_0}^2$.
\end{itemize}
For the bulge NEMO model which stands for our visible part of the galaxy we need 3 parameters:
\begin{itemize}
\item $\rho_b$~-- central density;
\item $\Psi_c$~-- bulge cut-off potential;
\item $\sigma_b$~-- bulge central potential.
\end{itemize}

The names of parameters for mkkd95 code
can be found in the table~\ref{NEMO_code} in the Appendix.

\begin{figure}
\caption{
Initial bulge density distribution for model~1
and different visible density profiles
see sec.~\ref{Nemo_modeling} for details.
\label{start_rhoLum_model1}
}
\includegraphics[width=0.45\textwidth]{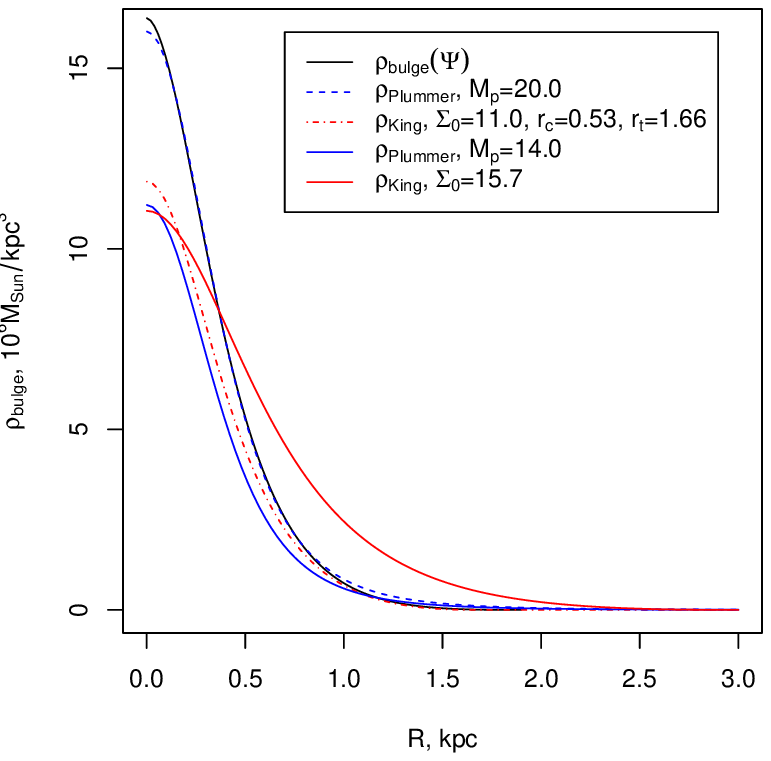}
\end{figure}

First we relied on the density distributions of
the components mentioned in the previous section
(the density distribution, eq.~(\ref{King_dist}), for the visible part,
and eq.~(\ref{rhoDM_distribution})
for the DM part of the galaxy) to calculate parameters
$v_0$, $R_a$, ${r_{ck}}^2$, $\rho_b$, $\Psi_c$, $\sigma_b$.
With this approach we failed to construct the NEMO-model by the mkkd95 code.
The next idea was to fit the density distributions (eqs.~\ref{KD95_eq2} and~\ref{KD95_eq4})
directly to the density distributions 
(eqs.~\ref{Plum_dist} and~\ref{rhoDM_distribution}).
To do this, we need an initial guess on the distribution 
of the potential derived from our density distributions.
We have calculated the combined potential distribution 
$\Psi(R,z=0)$ in cylindrical coordinates in the
equatorial plane $z=0$ (see the Appendix).

\begin{figure}
\caption{Initial halo density distribution for model~1
and~\citet{H16} DM density profile. 
\label{start_rhoDM_model1}
}
\includegraphics[width=0.45\textwidth]{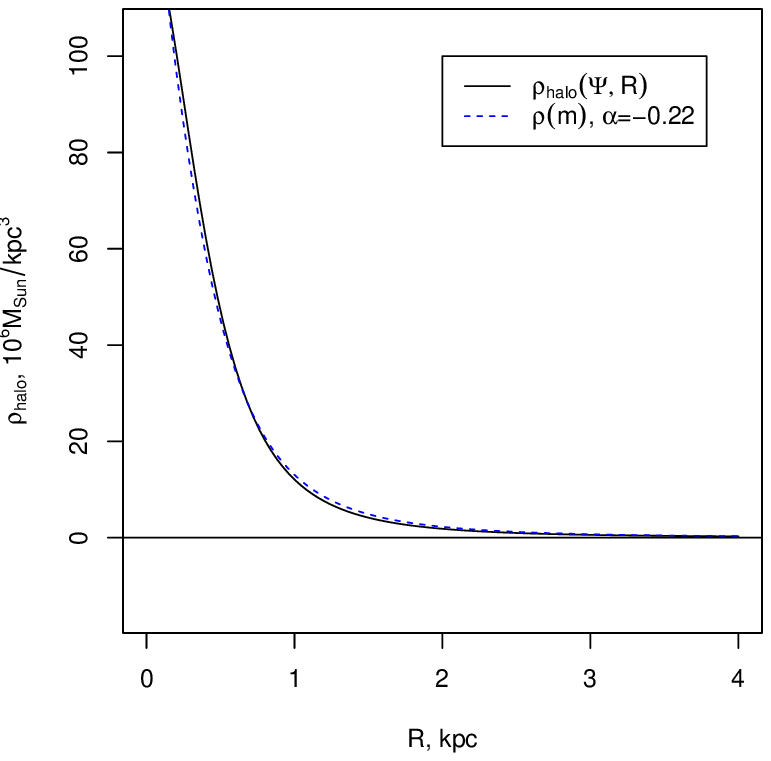}
\end{figure}

For the first model we need to find the central potential $\Psi_0$
as the sum of the central prolate DM potential
and the central stellar potential.
For calculating central DM potential we use eq.~(\ref{Phi_prolate})
for the case $r=0, z=0$.
As the $\mu$ function for this formula we used
eq.~(\ref{Psi_dep_p}) for the prolate cusped 
Zhao DM potential.
For the Plummer oblate central stellar potential
we used eq.~(\ref{Plummer_phi}).

After that we do the fitting procedure for the function eq.~(\ref{KD95_eq2})
(with known $\Psi_0$ and
three free parameters  $\Psi_c$, $\sigma_b$ and $\rho_b$)
to the Plummer density distribution function eq.~(\ref{Plum_dist}).
To do this we need to
numerically integrate eq.~(\ref{dPhi_prolate}), 
the same as eq.~(\ref{dPhi_Q})
as the growth of the DM potential over its central value.
We also need to get the limit of the
formula for the stellar eq.~(\ref{Plummer_phi})
radial potential distribution at the equatorial plane of the galaxy.
(We use the value $x=10000$ instead of $\infty$.)

\begin{deluxetable*}{cRRDcRRDRR}
\tablecaption{Parameters for NEMO models.\label{Nemo_params}}
\tablehead{
\colhead{Model} &\colhead{$\rho_b$}&\colhead{$\psi_c$}&\multicolumn2c{$\sigma_b$}&\colhead{$q$}&\colhead{$\Psi_0$}&\colhead{$v_0$}&\multicolumn2c{$R_a$}&\colhead{${r_{ck}}^2$}&\colhead{$\Delta$}
}
\decimals
\startdata
1                 &  20.75 & -277.5   & 7.796 &$1.11$ & -494.6 & 14.97 & 642.5 & 1.3*10^{11} & 11.8\%  
\enddata
\end{deluxetable*}

In Figure~\ref{start_rhoLum_model1}
the fitted function $\rho_{\rm bulge}(\Psi)$
from eq.~(\ref{KD95_eq2}),
shown by the black line, almost
coincides with the data curve $\rho_{\rm{Plummer}}$ 
with $M_p=20.0$ parameter
from eq.~(\ref{Plum_dist}), depicted by the blue dash-line.
This value of the parameter makes the stellar density
profile as the one taken by~\citet{H16}.
This profile was also assumed when constucting
the AGAMA model~3.
The red dash-dotted line represents the King profile
taken for the AGAMA model~2. 
This profile lies near the Plummer density profile 
shown by the blue solid line
with the stellar mass $\sim 20.0 \times 10^6 M_\odot$
from~\citet{LDwf} and $M_p=14.0$ parameter.
The stellar profile for model~2 lies also near the 
King profile given by~\citet{IH} shown 
by the red solid line.

The next step is fitting the function eq.~(\ref{KD95_eq4}) 
with three free parameters $\sigma_0$, $B$ and $C$ to 
the DM density distribution
function eq.~(\ref{rhoDM_distribution}).
In Figure~\ref{start_rhoDM_model1}
the fitted function $\rho_{\rm{halo}}(\Psi,R)$
from eq.~(\ref{KD95_eq4}),
shown by the black line,
almost coinsides with the DM density profile
$\rho(m)$ from~\citet{H16}
(see eq.~\ref{rhoDM_distribution}).
This curve is shown by the blue dash-line.
Then we get $v_0$ from $\sigma_0$ and $R_a$
value from those parameters using the expression in 
eq.~(\ref{r_a}).
To find the ${r_{ck}}^2$ value we need to calculate $R_c$ using
eqs.~(\ref{B},~\ref{C}) and $R_K$.
This value is the radius at which
 the rise of the combined potential over its central value
 is equal to $2{\sigma_0}^2$.

The fitting procedures were done for the density distributions at
the equatorial plane of the galaxy.
By these two fits
 we have gotten 5 parameters for the
  mkkd95 NEMO modeling
listed in Table~\ref{Nemo_params}.

The parameters taken from the fitting procedures 
(Table~\ref{Nemo_params})
are the initial parameters for the NEMO-modeling.
We can also see a variable $\Delta$ in Table~\ref{Nemo_params}
which is a measure of the violation of the virial theorem
 at the initial point of the numerical falcON evolution
 of the constructed galaxy model, namely,
$\Delta=-(2T/W)-1$, where $T$ is the kinetic energy
and $W < 0$ is the potential energy of the system.

\subsection{Orbit-based AGAMA modeling}
 The AGAMA framework developed by \citet{Vasiliev2018} is able to form an orbit-based model from some specific
and common patterns of densities and DFs.

For the DM halo we use the Spheroid AGAMA
component of the following form:
\begin{multline}
\label{DM_AGAMA}
\rho(\bar{r}) = \rho_0 \left( \frac{\bar{r}}{b_{\rm halo}} \right)^{\alpha}
  \left[ 1 + \left(\frac{\bar{r}}{b_{\rm halo}} \right)^{-\gamma}
   \right]^\frac{-\alpha-\eta}{\gamma} \times \\
 \times \exp \left[ - \left(\frac{\bar{r}}{r_{\rm cut}}\right)^\xi \right] \; ,
 \; \bar{r} = \sqrt{ x^2 + (y/p)^2 + (z/Q)^2 } \; .
\end{multline}
Far from the $r_{\rm cut}$ radius the above form coincides with the
DM profile (eq.~\ref{rhoDM_distribution}) when
\begin{multline}
\gamma = 2 \; , \; \eta = 3 \; , \; p=1.0 \; .
\end{multline}
For all models we set $\eta=3$, and $\gamma=2.0$ 
(see table~\ref{AGAMA_DM_params}).
As for the $r_{\rm cut}$ and $\xi$ parameters we use the following
numbers for all our AGAMA models:
\begin{eqnarray}
r_{\rm cut} = 55 , \; \xi =  2.5 \; .
\end{eqnarray}

The visible component is set either by the Plummer or 
King density profiles.
To construct the Plummer component using the
orbit-based AGAMA code we need the parameters
$Q, p, b_p$, and $M_{\rm stars}$ 
for the total component mass parameter.
Taking into account the formulas~\ref{Plum_oblate_mass} 
from the Appendix
we can estimate this mass from the $M_p$ parameter:
\begin{eqnarray}
M_{\rm stars} = q M_p .
\end{eqnarray}

For the King component we need
$M_{\rm stars}, r_{cA}$ and
$W_0 = [\Psi(r_t) - \Psi(0)]/\sigma^2$ parameters.
 The latter is the dimensionless potential depth of the generalized King
 (lowered isothermal) models (see AGAMA documentation).
The identifiers of parameters for the AGAMA Schwarzschild's
code can be found in Table~\ref{AGAMA_code} in the Appendix.

\begin{deluxetable}{cCCCCC}
\tablecaption{\bf Parameters for AGAMA DM component.\label{AGAMA_DM_params}}
\tablehead{
\colhead{Model} & \colhead{$\rho_0$}  & \colhead{$b_{\rm halo}$} & \colhead{$\gamma$} & \colhead{$\alpha$} & \colhead{$Q$}
}
\startdata
2              & 87.1               &          0.617         &      2.0         &      0.0         &   1.11 \\ 
3              & 87.1               &          0.617         &      2.0         &     -0.22        &   1.11 \\ 
\enddata
\end{deluxetable}

We have gotten two AGAMA models, one of them having
the cusped DM density profile as~\citet{H16}.

\begin{deluxetable}{cccccrc}
\tablecaption{Parameters for AGAMA visible component.\label{AGAMA_vis_params}}
\tablehead{
\colhead{Model}&\colhead{$M_{\rm stars}$}&\colhead{$r_{cA}$}       &\colhead{$W_0$}             &\colhead{$\beta$}&\colhead{$\beta_z$}&\colhead{$\kappa$}
}
\startdata
\multicolumn{7}{c}{King Visible component} \\
2              &              $12.13$    &    $0.753$           &        $1.786$           &      $-0.17$    &          -        &          -       \\ 
\multicolumn{7}{c}{Plummer visible component}\\
               &                         &    $b_p$             &         $q$              &                 &                   &                  \\
3              &              $13.19$    &    $0.668$           &       $0.66$             &           -     &       $0.47$      &        $1.0$     \\ 
\enddata
\end{deluxetable}

The first AGAMA model~2 assumes the visible density King profile with
the velocity anisotropy $\beta=-0.17$.
The characteristic radius 
and density for the halo component
are
from~\citet{H16} but the profile is cored.
The second AGAMA model~3 employs the Plummer visible density profile and more
appropriate velocity anisotropy parameters
for the model of~\citet{H16} and the same parameters
for the DM component.

\section{Results}

\begin{figure}[t]
\caption{\bf Evolution of the virial ratio}
One time unit corresponds to $0.47$ Gyr
(see eq.~\ref{time_unit}).
NEMO and AGAMA models are depicted all together.
AGAMA lines merge to one multicolored curve
after about 10 time units.
In contrast to NEMO model, there is no sense to
follow the AGAMA evolution further.
\label{virial_ratios} 
\includegraphics[height=0.4\textheight]{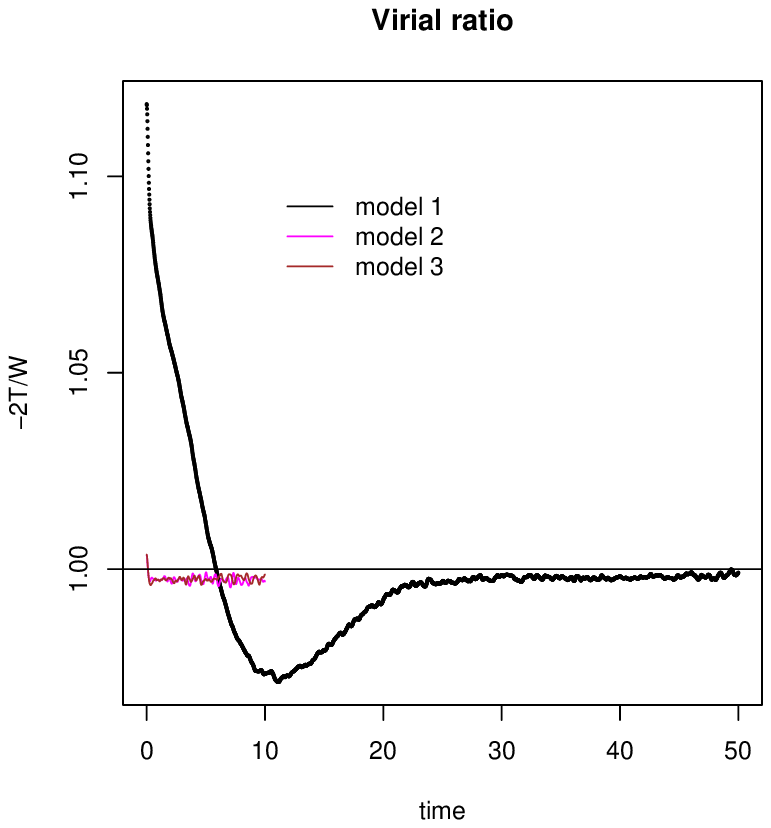}
\end{figure}

Figure~\ref{virial_ratios}
demonstrat that orbit-based AGAMA code produces the systems which are 
in much better equilibrium than the
ones from the mkkd95 NEMO code.
The falcON code computes the virial ratio for each step
of system evolution.
The values of this ratio  for
all our models over the time evolution steps
are depicted in Figure~\ref{virial_ratios}.
We can see that the AGAMA values $\Delta$ for the
violation of the virial theorem
(models~2--3) are negligible in comparison to the 
ones for the NEMO model.

From this figure, all of models
reach to the equilibrium sooner or later.

\subsection{NEMO modeling}

\begin{figure*}
\centering
  \begin{minipage}{7in}
\begin{tabular}{cc}
Stars projected surface density profile & DM projected surface density profile \\
\includegraphics[width=0.45\textwidth]{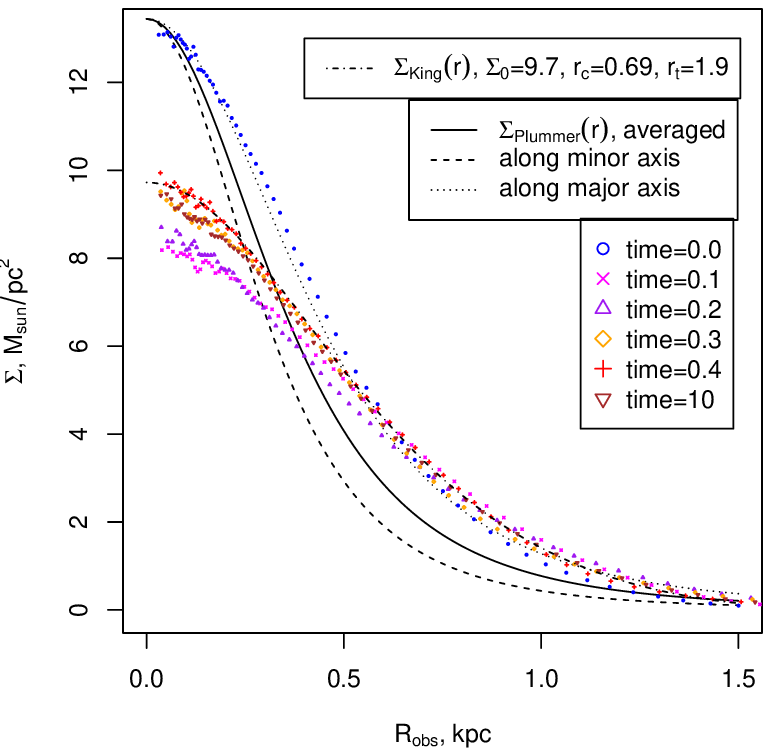}&%
\includegraphics[width=0.45\textwidth]{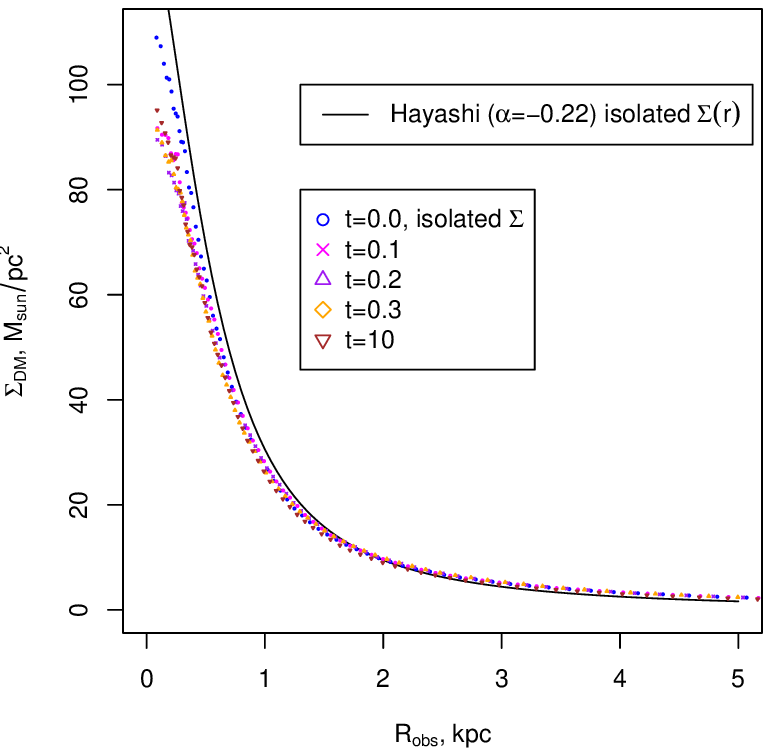}
\end{tabular}
\caption{Surface density evolution of components for NEMO model}
\label{Sigmas_evolution_nemo}
  \end{minipage}
\end{figure*}

\begin{figure*}
\includegraphics[height=0.24\textheight]{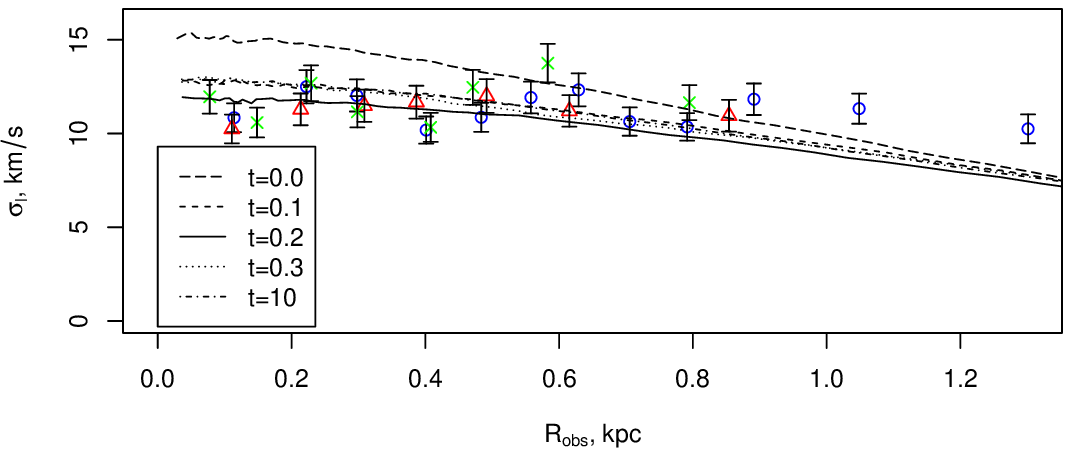}
\caption{NEMO model vs observed stars velocity dispersion profile}
\label{Velocity_dispersions_nemo}
The blue circles represent data measured along the
major galaxy axis,
the red triangles represent data measured along the minor axis,
and the green crosses represent data measured
along the middle axis.
\end{figure*}

The~\citet{Den} manipulators for the N-body galaxy snapshot files
are able to calculate the profiles
of different physical quantities averaged either over
the spherical radius or over the cylindrical radius.
In the latter case we use the projection along the line-of-sight.
Figure~\ref{Sigmas_evolution_nemo}
shows the evolution of
modeled surface density profiles
of the visible and DM components and the initial analytical profiles.
There are several analytical profiles of
the surface density along different axes and the averaged surface density.

We also found the best-fit parameters of the King distribution
for time $t=0.4$ modelled profile.
The fitting dash-dotted line is hidden by orange, red and brown symbols
standing for $t=0.3$, $t=0.4$ and $t=10$ density profiles.
The parameters for the line are included in the legend
of Figure~\ref{Sigmas_evolution_nemo}.

For the model we can see that the stellar surface density profile 
(in the left panel of Figure 4)
falls down by about 20\% at the first evolution steps 
and then fluctuates about its equilibrium position.
This falling is a factor of two faster than the falling of the
other our NEMO models not presented here. But this model shows the best
 agreement with the velosity dispersion data among all our NEMO models.

Earlier than $t \sim 10.0$  the virial ratio does not reach
its equilibrium value (Figure~\ref{virial_ratios}) but
but we can say that all further fluctuations
of projected profiles lie between $t=0.1$ and $t=0.4$ lines. 

In the right panels of Figure~\ref{Sigmas_evolution_nemo}, 
we can see smaller relative drop
of the surface density during the first evolution steps in comparison
with the visible density profile.

We can compare modelled velocity dispersion projected onto the line-of-sight
and averaged over the cylindrical radius of the sky plane profile
and its falcON evolution directly
to the observational velocity dispersion points.
We need to multiply projprof Dehnen manipulator velocity data by
$V_{\rm scale} = 2.079$ in order to convert $\sqrt{10^6 M_\odot / \mbox{kpc}}$
velocity units to $\mbox{km/s}$.

We use $\chi^2$ test to compare our models:
\begin{multline}
\chi^2 = \sum_{i=1}^N \frac{(\sigma_l({R_{\rm obs}}^i) - ({\sigma_{\rm obs}}^i))^2}{{({\delta\sigma_{\rm obs}}^i)}^2} \; ,
\end{multline}
where $\sigma_l({R_{\rm obs}}^i)$ are mean modelled
velocity dispersions projected to the line-of-sight
over the modelled star points spaced ${R_{\rm obs}}^i$ apart the line-of-sight axe,
${\sigma_{\rm obs}}^i$ are the observational velocity dispersions
at ${R_{\rm obs}}^i$ radii
and ${\delta \sigma_{\rm obs}}^i$ are observational errors.

We do not perform the fitting procedure for minimizing $\chi^2$
by varying parameters because each set of the parameters
stands for one $~10^6$ particles N-body simulation
so fitting procedure would take too many computational resources.
That is why we use hydrodynamical parameters to start our modeling.
The mean value of
$\overline{\chi^2}$ averaged over time for $t\ge 0.35$
is $83.5$ for model~1.
In the Appendix~\ref{NEMO_detailed} we discuss
the time interval for averaging and the equilibrium velocity
dispersion profile which becomes so much earlier than the
$\Delta$ parameter does.
Figure~\ref{Velocity_dispersions_nemo} represents the
velocity dispersion projected to the line-of-sight
profile for several evolution time points of the
falcON in comparison to the observational data with error bars.

\subsection{AGAMA modeling}

\begin{figure*}
\centering
  \begin{minipage}{7in}
\begin{tabular}{ccc}
& Model~2 & Model~3 \\
\rotatebox[y=0.25\textwidth]{90}{Stars projected surface density}&%
{\includegraphics[width=0.45\textwidth]{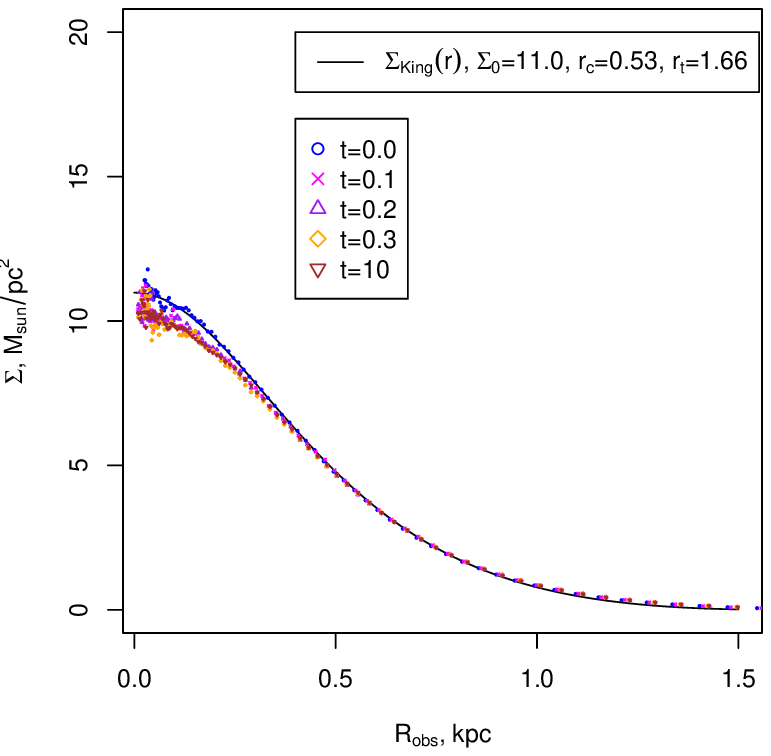}}&%
{\includegraphics[width=0.45\textwidth]{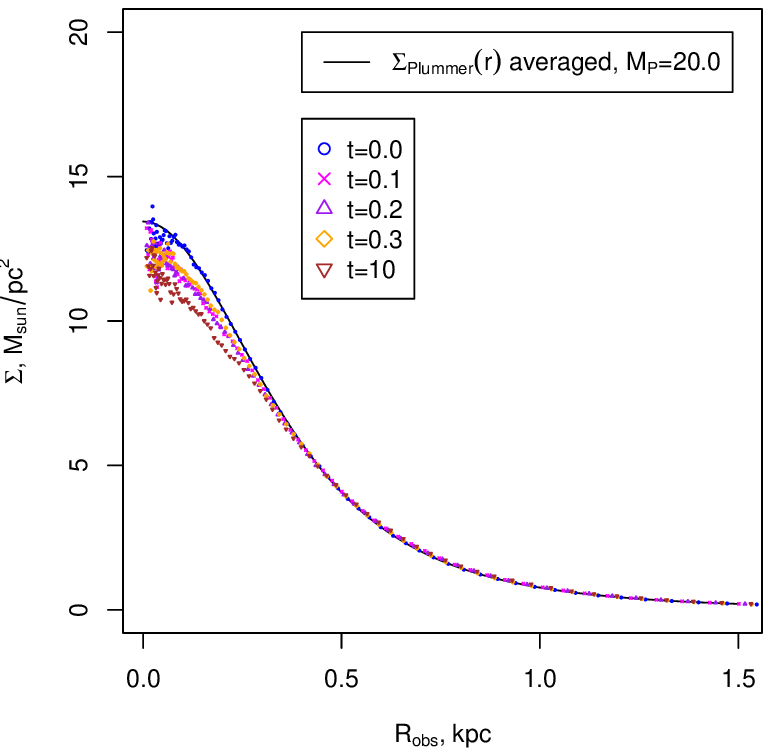}}\\
\rotatebox[y=0.25\textwidth]{90}{DM projected surface density}&%
{\includegraphics[width=0.45\textwidth]{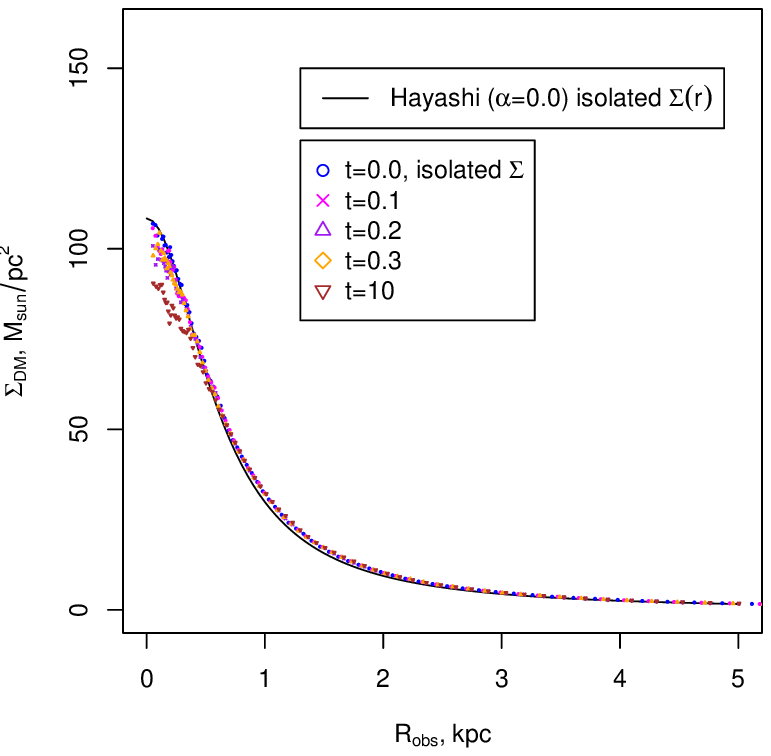}}&%
{\includegraphics[width=0.45\textwidth]{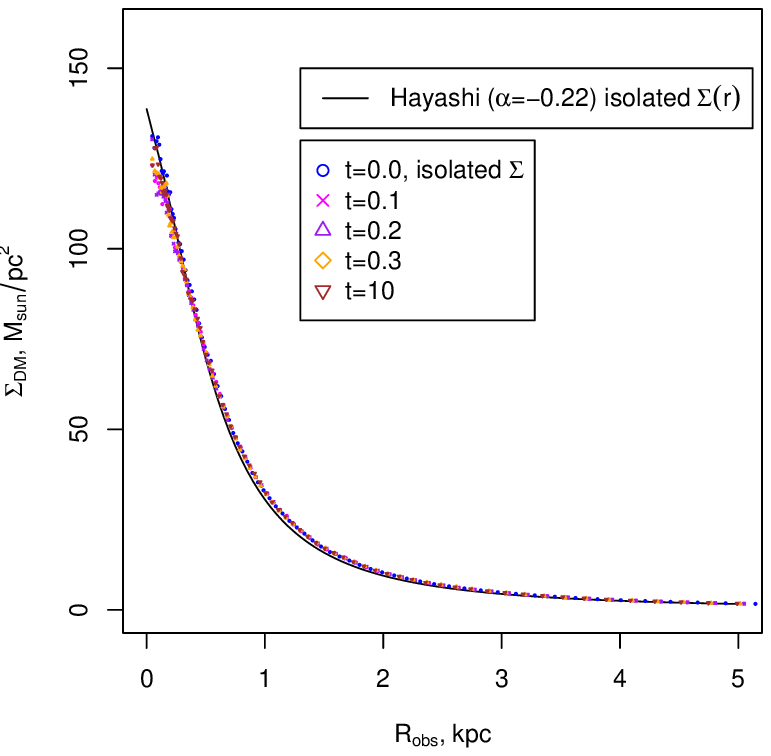}}
\end{tabular}
\caption{Surface density evolution of components for AGAMA models}
\label{Sigmas_evolution_AGAMA}
  \end{minipage}
\end{figure*}

The AGAMA modeled visible and DM surface density profiles are depicted
in the figure~\ref{Sigmas_evolution_AGAMA}.
We can see the decrease of stellar surface density profile for the
equilibrium establishing time interval. And such decreasing in the
central part of the galaxy is less than that for the
NEMO model~1.

But in contrast to
this model the changing of density profiles over time for the AGAMA systems
is noticeable only at the central part of the galaxy
with $r\lesssim 0.25\mbox{ kpc}$.
For the NEMO model the radius
of noticeable change of the surface density profile is not less than
$0.5\mbox{ kpc}$.

For model~2 we found $\Sigma_0, r_c, r_t$
King profile parameters
that fit well the start surface density profile with the initial
$W_0, r_c$ and $M_{\rm stars}$
AGAMA parameters.
They are shown on the panel for this model in the figure~\ref{Sigmas_evolution_AGAMA}.

For model~3 we depicted the averaged Plummer
surface density 
profile as a black line
taking into account our start parameters.

The second row of Figure~\ref{Sigmas_evolution_AGAMA} is shown by
the DM surface density profiles. All the initial profiles 
almost coincide with the
analytical profiles taking our initial parameters.
We can also see small evolution changes of these profiles. 
The greatest change for model~2 is less
than the smallest one for our NEMO model.

Model~3 has almost the same initial parameters as the
NEMO-model~1 which coincides with the~\citet{H16} parameters.
Model 3 changes less than the
corresponding NEMO-model but the ensuing evolution curves of the model~1
intersect more data error bars than those of the AGAMA model.

\begin{figure*}
\begin{tabular}{cc}
\rotatebox{90}{Model~2}&\parbox{0.8\textwidth}{\includegraphics[height=0.24\textheight]{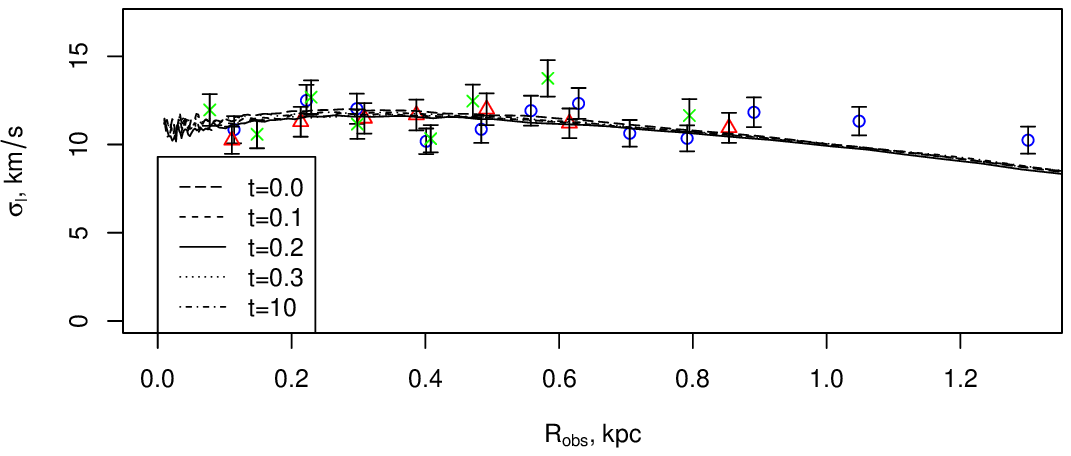}}\\
\rotatebox{90}{Model~3}&\parbox{0.8\textwidth}{\includegraphics[height=0.24\textheight]{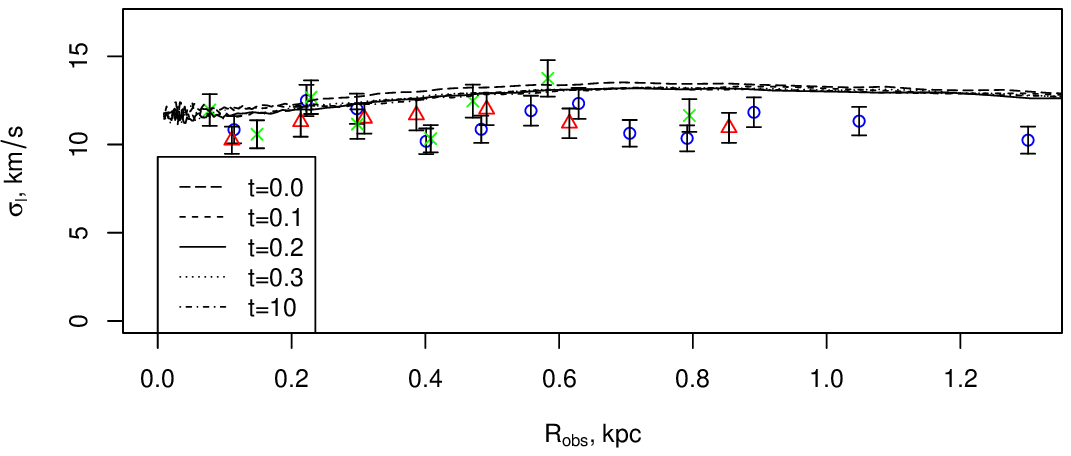}}\\
\end{tabular}
\caption{AGAMA {\bf models} vs observed stars velocity dispersion profiles}
\label{Velocity_dispersions_AGAMA}
\end{figure*}

\begin{deluxetable}{cDDD}
\tablecaption{Mean values of $\chi^2$ \label{Chi_sq}}
\tablehead{
\colhead{}         &\multicolumn2c{1}&\multicolumn2c{2}&\multicolumn2c{3}
}
\decimals
\startdata
$\overline{\chi^2}$&         83.5    &         36.5    &     104.7
\enddata
\end{deluxetable}

We can see the averaged
values of $\chi^2$ over time for $t\ge 0.2$ for
the AGAMA models and for $t\ge 0.35$ for
the NEMO model
in Table~\ref{Chi_sq}.
In the Appendix~\ref{AGAMA_detailed} the discussion about
this averaging can be found.
The mean equilibrium $\chi^2$ value for the NEMO model~1
is less than that of the AGAMA model~3.

The worst $\chi^2$ is for
model~3 with $\alpha=-0.22$
and $\beta_z=0.47$.
It seems that this galaxy shows better cored DM profile.
Among the multitude of our other models we can say that exact~\citet{H16}
profile is better with $\beta_z=0.47$ velocity anisotropy parameter
and $\alpha=0.0$ or $\gamma=1.0$ DM profiles are better with
$\beta_z=-0.17$ or $\beta=-0.17$ parameters.

In Figure~\ref{Velocity_dispersions_AGAMA} we can see that the
AGAMA models are compliant with the observational data
with different form of curves.
Notation of the points
is the same as in Figure~\ref{Velocity_dispersions_nemo}.

\begin{deluxetable}{cCCC}
\tablecaption{King best-fit parameters 
vs obervations.\label{King_fit}}
\tablehead{
&\colhead{$\Sigma_0, L_\odot/\mbox{pc}^2$}&\colhead{$r_c,\mbox{ kpc}$}&\colhead{$r_t,\mbox{ kpc}$}
}
\startdata
observed 1               & 15.7 \pm 5.1   &  0.75 \pm 0.01  &  2.95 \pm 0.02                    \\
observed 2               & 15.7 \pm 5.1   &  0.58 \pm 0.05  &  3.04 \pm 0.17                    \\
model~1                  &                       9.7        &  0.69           &  1.9            \\
model~2                  &                       11.0       &  0.53           &  1.66           \\
\enddata
\tablecomments{
Fitted to observation structural parameters from
\citet{Bat}~-- observed 1;  \citet{IH}~-- observed 2.
}
\end{deluxetable}

Let us compare the best NEMO and AGAMA models
stellar component equilibrium profiles.
The King parameters for them are listed in Table~\ref{King_fit}.
We can see that for both models $\Sigma_0$ is on the bottom edge
of the observed error bar or a bit lower.
The structural parameters are closer to
the~\citet{Bat} data.

\section{Comparison with previous work}
Let us compare our results with the investigations
of~\citet{Pascale}, \citet{Amorisco} and~\citet{Strigari}.
We can estimate the dynamical mass $M_{\rm dyn}(r)$ of our best model inside the
spheres of $1.05$ kpc and $300$ pc radii.

For $M_{\rm dyn}(1.05\mbox{ kpc})$ we have $1.29\times 10^8 M_\odot$
for our model~2. The estimation of~\citet{Pascale} for this mass
is $1.38\pm 0.10 \times 10^8 M_\odot$ compared with
the estimation of~\citet{Amorisco} at $1.3 \times 10^8 M_\odot$.
We can see that our numbers are consistent with the previous results.

The mass enclosed in a sphere of $300$ pc radius is
$0.93 \times 10^7 M_\odot$ for model~2.
The results of~\citet{Strigari} for $M_{\rm dyn}(0.3\mbox{kpc})$ is
$1.14^{+0.09}_{-0.12} \times 10^7 M_\odot$.
Smaller value at $0.44^{+0.07}_{-0.03} \times 10^7 M_\odot$
is the estimation of~\citet{Pascale}. We can see that our
numbers for a smaller radius is closer to the older estimations
of~\citet{Strigari}.

We can also compare our results to the DM halo density
at $150\mbox{ pc}$ radius modeled by~\citet{Read2019}
by their GravSphere code based on the spherical Jeans
equations and their sophisticated assumptions about
the {\bf velocity} anisotropy parameter. They obtained the value 
$\rho_{\rm DM}(150\mbox{ pc}) = 0.79^{+0.27}_{-0.19} \times 10^8 M_\odot/\mbox{kpc}^3$.
The values of this density for all our models are inside
this error bar. We can say that our results are also
{\bf consistent} with the estimations of~\citet{Read2019}.

\section{Conclusion}
The main point of this paper is that we first investigate 
whether the dynamical non-spherical structure of Fornax dSph 
estimated from kinematic analysis by~\citep{H16} is stable or not.
We have constructed the N-body models by two source codes for the 
Fornax galaxy from these hydrodynamical studies
and followed their numerical evolution. 

Then we found that the AGAMA model with the 
inferred dark halo parameters
demonstrates that indeed, such a
system may be stable during several dynamical times.

Both methods (NEMO and AGAMA) used in our paper give approximate
realizations of stationary models. AGAMA is much more accurate, which is
clear from the criterion of virial test. Moreover, AGAMA can reproduce the
non-spherical self-consistent structure of Fornax
as the weighted sum of orbit contributions to the galactic density
even though the inferred dark halo 
parameters come from Jeans analysis which does not require 
that any distribution functions should be positive.

We have also changed some assumptions about the stellar and DM density
distributions and got the model that fits the kinematical data
much better than the models found in previuos studies.
Better DM profiles are cored profiles.
The velocity anisotropy parameter for the stellar component
 also shows some correlations with
the DM profile characteristics.
The best stellar profile is the King profile.
We choose this model in favour of the best agreement with the velocity dispersion data
than with the luminosity data because the velocity dispersion observations are
more precise than the luminosity ones.

One should understand that even
exactly stationary DF-based models may be dynamically or secularly unstable
(there are numerous well-known stationary solutions for collisionless
disk-like or ellipsoidal systems which may be unstable).
Thus, N-body evolution is needed to test not only the deviations from
the exact stationary solution but also to check the stability.
Our results provided by falcON simulations do show that all our models
are dynamically stable since they support their shapes with small
oscillations for several dynamical time-scales.

Using N-body simulation results, one  can confine the prior
distributions of dark matter halo parameters such as scale radius and
scale density.
Based on observational results \citep[such as Jeans analysis by][]{H16} our analysis,
applied  to other dwarf galaxies,
can obtain the limits on the parameter ranges of
the dark matter halo which sustain dynamical equilibrium.
Therefore, using the confined parameter ranges, we may improve the
limits on the nature of DM particle through its annihilation~\citep{2020arXiv200211956A}.

\section*{Acknowledgements}

The authors are grateful to Eugene Vasiliev for his invaluable help with the AGAMA code
and to Marat Potashov for debugging FalcOn code.

We are grateful to the anonymous referee for
useful comments which helped us to improve the presentation
of our results.

This work was supported by JSPS KAKENHI Grant Numbers,
18H04359 \& 18J00277 for K.H.

S.B. is grateful to Sternberg Astronomical Institute, MSU, for generous support.

\bibliography{NbodySHmodel_arXiv}{}

\begin{thebibliography}{}
\expandafter\ifx\csname natexlab\endcsname\relax\def\natexlab#1{#1}\fi

\bibitem[{{Amorisco} \& {Evans}(2011)}]{Amorisco}
{Amorisco}, N.~C., \& {Evans}, N.~W. 2011, \mnras, 411, 2118

\bibitem[{{Ando} {et~al.}(2020){Ando}, {Geringer-Sameth}, {Hiroshima}, {Hoof},
  {Trotta}, \& {Walker}}]{2020arXiv200211956A}
{Ando}, S., {Geringer-Sameth}, A., {Hiroshima}, N., {et~al.} 2020, \prd, 102,
  061302

\bibitem[{{Babcock}(1939)}]{Babcock1939}
{Babcock}, H.~W. 1939, Lick Observatory Bulletin, 498, 41

\bibitem[{{Battaglia} {et~al.}(2013){Battaglia}, {Helmi}, \&
  {Breddels}}]{Battaglia13}
{Battaglia}, G., {Helmi}, A., \& {Breddels}, M. 2013, \nar, 57, 52

\bibitem[{{Battaglia} {et~al.}(2006){Battaglia}, {Tolstoy}, {Helmi}, {Irwin},
  {Letarte}, {Jablonka}, {Hill}, {Venn}, {Shetrone}, {Arimoto}, {Primas},
  {Kaufer}, {Francois}, {Szeifert}, {Abel}, \& {Sadakane}}]{Bat}
{Battaglia}, G., {Tolstoy}, E., {Helmi}, A., {et~al.} 2006, \aap, 459, 423

\bibitem[{{Bekenstein}(2004)}]{Bekenstein2004}
{Bekenstein}, J.~D. 2004, \prd, 70, 083509

\bibitem[{{Bekenstein}(2010)}]{Bekenstein2010}
---. 2010, {Modified gravity as an alternative to dark matter}, ed.
  G.~{Bertone}, 99

\bibitem[{{Binney} \& {Tremaine}(2008)}]{BT}
{Binney}, J., \& {Tremaine}, S. 2008, {Galactic Dynamics: Second Edition}
  (Princeton University Press)

\bibitem[{{Breddels} \& {Helmi}(2013)}]{BreddelsHelmi13}
{Breddels}, M.~A., \& {Helmi}, A. 2013, \aap, 558, A35

\bibitem[{{Casas} {et~al.}(2017){Casas}, {Kunz}, {Martinelli}, \&
  {Pettorino}}]{MG2017}
{Casas}, S., {Kunz}, M., {Martinelli}, M., \& {Pettorino}, V. 2017, Physics of
  the Dark Universe, 18, 73

\bibitem[{{de Blok} {et~al.}(2001){de Blok}, {McGaugh}, {Bosma}, \&
  {Rubin}}]{deBlok2001}
{de Blok}, W.~J.~G., {McGaugh}, S.~S., {Bosma}, A., \& {Rubin}, V.~C. 2001,
  \apj, 552, L23

\bibitem[{{Dehnen}(2002)}]{Den}
{Dehnen}, W. 2002, Journal of Computational Physics, 179, 27

\bibitem[{{Einasto} {et~al.}(1974){Einasto}, {Kaasik}, \& {Saar}}]{Einasto1974}
{Einasto}, J., {Kaasik}, A., \& {Saar}, E. 1974, \nat, 250, 309

\bibitem[{{Evans} \& {Collett}(1993)}]{Evans93}
{Evans}, N.~W., \& {Collett}, J.~L. 1993, \mnras, 264, 353

\bibitem[{{Gilmore} {et~al.}(2007){Gilmore}, {Wilkinson}, {Wyse}, {Kleyna},
  {Koch}, {Evans}, \& {Grebel}}]{Gilmore2007}
{Gilmore}, G., {Wilkinson}, M.~I., {Wyse}, R. F.~G., {et~al.} 2007, \apj, 663,
  948

\bibitem[{{Gonz{\'a}lez-Morales} {et~al.}(2017){Gonz{\'a}lez-Morales}, {Marsh},
  {Pe{\~n}arrubia}, \& {Ure{\~n}a-L{\'o}pez}}]{Gonzalez-Morales}
{Gonz{\'a}lez-Morales}, A.~X., {Marsh}, D. J.~E., {Pe{\~n}arrubia}, J., \&
  {Ure{\~n}a-L{\'o}pez}, L.~A. 2017, \mnras, 472, 1346

\bibitem[{{Hayashi} \& {Chiba}(2012)}]{HC12}
{Hayashi}, K., \& {Chiba}, M. 2012, \apj, 755, 145

\bibitem[{{Hayashi} \& {Chiba}(2015)}]{HC15}
---. 2015, \apj, 810, 22

\bibitem[{{Hayashi} {et~al.}(2016){Hayashi}, {Ichikawa}, {Matsumoto}, {Ibe},
  {Ishigaki}, \& {Sugai}}]{H16}
{Hayashi}, K., {Ichikawa}, K., {Matsumoto}, S., {et~al.} 2016, \mnras, 461,
  2914

\bibitem[{{Irwin} \& {Hatzidimitriou}(1995)}]{IH}
{Irwin}, M., \& {Hatzidimitriou}, D. 1995, \mnras, 277, 1354

\bibitem[{{Jardel} \& {Gebhardt}(2012)}]{JG12}
{Jardel}, J.~R., \& {Gebhardt}, K. 2012, \apj, 746, 89

\bibitem[{{Jeffreson} {et~al.}(2017){Jeffreson}, {Sanders}, {Evans},
  {Williams}, {Gilmore}, {Bayo}, {Bragaglia}, {Casey}, {Flaccomio},
  {Franciosini}, {Hourihane}, {Jackson}, {Jeffries}, {Jofr{\'e}}, {Koposov},
  {Lardo}, {Lewis}, {Magrini}, {Morbidelli}, {Pancino}, {Randich}, {Sacco},
  {Worley}, \& {Zaggia}}]{Jeffreson17}
{Jeffreson}, S.~M.~R., {Sanders}, J.~L., {Evans}, N.~W., {et~al.} 2017, \mnras,
  469, 4740

\bibitem[{{King}(1962)}]{K62}
{King}, I. 1962, \aj, 67, 471

\bibitem[{{Klypin} \& {Shandarin}(1983)}]{1983MNRAS.204..891K}
{Klypin}, A.~A., \& {Shandarin}, S.~F. 1983, \mnras, 204, 891

\bibitem[{{Kowalczyk} {et~al.}(2019){Kowalczyk}, {del Pino}, {{\L}okas}, \&
  {Valluri}}]{Kowalczyk19}
{Kowalczyk}, K., {del Pino}, A., {{\L}okas}, E.~L., \& {Valluri}, M. 2019,
  \mnras, 482, 5241

\bibitem[{{Kuijken} \& {Dubinski}(1994)}]{KD94}
{Kuijken}, K., \& {Dubinski}, J. 1994, \mnras, 269, 13

\bibitem[{{Kuijken} \& {Dubinski}(1995)}]{KD95}
---. 1995, \mnras, 277, 1341

\bibitem[{{Lin} \& {Faber}(1983)}]{Lin}
{Lin}, D.~N.~C., \& {Faber}, S.~M. 1983, \apjl, 266, L21

\bibitem[{{Mateo}(1998)}]{Mateo1998}
{Mateo}, M.~L. 1998, Annual Review of Astronomy and Astrophysics, 36, 435

\bibitem[{{McConnachie}(2012)}]{LDwf}
{McConnachie}, A.~W. 2012, \aj, 144, 4

\bibitem[{{Mestel}(1963)}]{Mestel1963}
{Mestel}, L. 1963, \mnras, 126, 553

\bibitem[{{Milgrom}(1983)}]{Milgrom1983}
{Milgrom}, M. 1983, \apj, 270, 365

\bibitem[{{Ostriker} {et~al.}(1974){Ostriker}, {Peebles}, \&
  {Yahil}}]{Ostriker1974}
{Ostriker}, J.~P., {Peebles}, P.~J.~E., \& {Yahil}, A. 1974, \apj, 193, L1

\bibitem[{{Pascale} {et~al.}(2018){Pascale}, {Posti}, {Nipoti}, \&
  {Binney}}]{Pascale}
{Pascale}, R., {Posti}, L., {Nipoti}, C., \& {Binney}, J. 2018, \mnras, 480,
  927

\bibitem[{{Pe{\~n}arrubia} {et~al.}(2008){Pe{\~n}arrubia}, {McConnachie}, \&
  {Navarro}}]{Penarrubia2008}
{Pe{\~n}arrubia}, J., {McConnachie}, A.~W., \& {Navarro}, J.~F. 2008, \apj,
  672, 904

\bibitem[{{Pietrzy{\'n}ski} {et~al.}(2009){Pietrzy{\'n}ski}, {G{\'o}rski},
  {Gieren}, {Ivanov}, {Bresolin}, \& {Kudritzki}}]{Piet}
{Pietrzy{\'n}ski}, G., {G{\'o}rski}, M., {Gieren}, W., {et~al.} 2009, \aj, 138,
  459

\bibitem[{{Planck Collaboration} {et~al.}(2014){Planck Collaboration}, {Ade},
  {Aghanim}, {Armitage-Caplan}, {Arnaud}, {Ashdown}, {Atrio-Barandela},
  {Aumont}, {Baccigalupi}, {Banday}, {Barreiro}, {Bartlett}, {Battaner},
  {Benabed}, {Beno{\^\i}t}, {Benoit-L{\'e}vy}, {Bernard}, {Bersanelli},
  {Bielewicz}, {Bobin}, {Bock}, {Bonaldi}, {Bond}, {Borrill}, {Bouchet},
  {Bridges}, {Bucher}, {Burigana}, {Butler}, {Calabrese}, {Cappellini},
  {Cardoso}, {Catalano}, {Challinor}, {Chamballu}, {Chary}, {Chen}, {Chiang},
  {Chiang}, {Christensen}, {Church}, {Clements}, {Colombi}, {Colombo},
  {Couchot}, {Coulais}, {Crill}, {Curto}, {Cuttaia}, {Danese}, {Davies},
  {Davis}, {de Bernardis}, {de Rosa}, {de Zotti}, {Delabrouille}, {Delouis},
  {D{\'e}sert}, {Dickinson}, {Diego}, {Dolag}, {Dole}, {Donzelli}, {Dor{\'e}},
  {Douspis}, {Dunkley}, {Dupac}, {Efstathiou}, {Elsner}, {En{\ss}lin},
  {Eriksen}, {Finelli}, {Forni}, {Frailis}, {Fraisse}, {Franceschi}, {Gaier},
  {Galeotta}, {Galli}, {Ganga}, {Giard}, {Giardino}, {Giraud-H{\'e}raud},
  {Gjerl{\o}w}, {Gonz{\'a }lez-Nuevo}, {G{\'o}rski}, {Gratton}, {Gregorio},
  {Gruppuso}, {Gudmundsson}, {Haissinski}, {Hamann}, {Hansen}, {Hanson},
  {Harrison}, {Henrot-Versill{\'e}}, {Hern{\'a}ndez-Monteagudo}, {Herranz},
  {Hildebrandt}, {Hivon}, {Hobson}, {Holmes}, {Hornstrup}, {Hou}, {Hovest},
  {Huffenberger}, {Jaffe}, {Jaffe}, {Jewell}, {Jones}, {Juvela},
  {Keih{\"a}nen}, {Keskitalo}, {Kisner}, {Kneissl}, {Knoche}, {Knox}, {Kunz},
  {Kurki-Suonio}, {Lagache}, {L{\"a}hteenm{\"a}ki}, {Lamarre}, {Lasenby},
  {Lattanzi}, {Laureijs}, {Lawrence}, {Leach}, {Leahy}, {Leonardi},
  {Le{\'o}n-Tavares}, {Lesgourgues}, {Lewis}, {Liguori}, {Lilje},
  {Linden-V{\o}rnle}, {L{\'o}pez-Caniego}, {Lubin}, {Mac{\'\i}as-P{\'e}rez},
  {Maffei}, {Maino}, {Mandolesi}, {Maris}, {Marshall}, {Martin},
  {Mart{\'\i}nez- Gonz{\'a}lez}, {Masi}, {Massardi}, {Matarrese}, {Matthai},
  {Mazzotta}, {Meinhold}, {Melchiorri}, {Melin}, {Mendes}, {Menegoni},
  {Mennella}, {Migliaccio}, {Millea}, {Mitra}, {Miville- Desch{\^e}nes},
  {Moneti}, {Montier}, {Morgante}, {Mortlock}, {Moss}, {Munshi}, {Murphy},
  {Naselsky}, {Nati}, {Natoli}, {Netterfield}, {N{\o}rgaard-Nielsen},
  {Noviello}, {Novikov}, {Novikov}, {O'Dwyer}, {Osborne}, {Oxborrow}, {Paci},
  {Pagano}, {Pajot}, {Paladini}, {Paoletti}, {Partridge}, {Pasian},
  {Patanchon}, {Pearson}, {Pearson}, {Peiris}, {Perdereau}, {Perotto},
  {Perrotta}, {Pettorino}, {Piacentini}, {Piat}, {Pierpaoli}, {Pietrobon},
  {Plaszczynski}, {Platania}, {Pointecouteau}, {Polenta}, {Ponthieu}, {Popa},
  {Poutanen}, {Pratt}, {Pr{\'e}zeau}, {Prunet}, {Puget}, {Rachen}, {Reach},
  {Rebolo}, {Reinecke}, {Remazeilles}, {Renault}, {Ricciardi}, {Riller},
  {Ristorcelli}, {Rocha}, {Rosset}, {Roudier}, {Rowan-Robinson},
  {Rubi{\~n}o-Mart{\'\i}n}, {Rusholme}, {Sandri}, {Santos}, {Savelainen},
  {Savini}, {Scott}, {Seiffert}, {Shellard}, {Spencer}, {Starck}, {Stolyarov},
  {Stompor}, {Sudiwala}, {Sunyaev}, {Sureau}, {Sutton}, {Suur-Uski}, {Sygnet},
  {Tauber}, {Tavagnacco}, {Terenzi}, {Toffolatti}, {Tomasi}, {Tristram},
  {Tucci}, {Tuovinen}, {T{\"u}rler}, {Umana}, {Valenziano}, {Valiviita}, {Van
  Tent}, {Vielva}, {Villa}, {Vittorio}, {Wade}, {Wandelt}, {Wehus}, {White},
  {White}, {Wilkinson}, {Yvon}, {Zacchei}, \& {Zonca}}]{Planck2014}
{Planck Collaboration}, {Ade}, P.~A.~R., {Aghanim}, N., {et~al.} 2014, \aap,
  571, A16

\bibitem[{{Plummer}(1911)}]{Plum}
{Plummer}, H.~C. 1911, \mnras, 71, 460

\bibitem[{{Read} {et~al.}(2019){Read}, {Walker}, \& {Steger}}]{Read2019}
{Read}, J.~I., {Walker}, M.~G., \& {Steger}, P. 2019, \mnras, 484, 1401

\bibitem[{{Richstone} \& {Tremaine}(1984)}]{Richstone}
{Richstone}, D.~O., \& {Tremaine}, S. 1984, \apj, 286, 27

\bibitem[{{Riess} {et~al.}(1998){Riess}, {Filippenko}, {Challis},
  {Clocchiatti}, {Diercks}, {Garnavich}, {Gilliland}, {Hogan}, {Jha},
  {Kirshner}, {Leibundgut}, {Phillips}, {Reiss}, {Schmidt}, {Schommer},
  {Smith}, {Spyromilio}, {Stubbs}, {Suntzeff}, \& {Tonry}}]{accelerationSNIa}
{Riess}, A.~G., {Filippenko}, A.~V., {Challis}, P., {et~al.} 1998, \aj, 116,
  1009

\bibitem[{{Safarzadeh} \& {Spergel}(2020)}]{Safarzadeh}
{Safarzadeh}, M., \& {Spergel}, D.~N. 2020, \apj, 893, 21

\bibitem[{{Salucci} {et~al.}(2012){Salucci}, {Wilkinson}, {Walker}, {Gilmore},
  {Grebel}, {Koch}, {Frigerio Martins}, \& {Wyse}}]{Salucci2012}
{Salucci}, P., {Wilkinson}, M.~I., {Walker}, M.~G., {et~al.} 2012, \mnras, 420,
  2034

\bibitem[{{Sanders} \& {Binney}(2016)}]{Sanders16}
{Sanders}, J.~L., \& {Binney}, J. 2016, \mnras, 457, 2107

\bibitem[{{Schwarzschild}(1979)}]{Schwarzschild}
{Schwarzschild}, M. 1979, \apj, 232, 236

\bibitem[{{Strigari} {et~al.}(2008){Strigari}, {Bullock}, {Kaplinghat},
  {Simon}, {Geha}, {Willman}, \& {Walker}}]{Strigari}
{Strigari}, L.~E., {Bullock}, J.~S., {Kaplinghat}, M., {et~al.} 2008, \nat,
  454, 1096

\bibitem[{{Tegmark} {et~al.}(2004){Tegmark}, {Blanton}, {Strauss}, {Hoyle},
  {Schlegel}, {Scoccimarro}, {Vogeley}, {Weinberg}, {Zehavi}, {Berlind},
  {Budavari}, {Connolly}, {Eisenstein}, {Finkbeiner}, {Frieman}, {Gunn},
  {Hamilton}, {Hui}, {Jain}, {Johnston}, {Kent}, {Lin}, {Nakajima}, {Nichol},
  {Ostriker}, {Pope}, {Scranton}, {Seljak}, {Sheth}, {Stebbins}, {Szalay},
  {Szapudi}, {Verde}, {Xu}, {Annis}, {Bahcall}, {Brinkmann}, {Burles},
  {Castander}, {Csabai}, {Loveday}, {Doi}, {Fukugita}, {Gott}, {Hennessy},
  {Hogg}, {Ivezi{\'c}}, {Knapp}, {Lamb}, {Lee}, {Lupton}, {McKay}, {Kunszt},
  {Munn}, {O'Connell}, {Peoples}, {Pier}, {Richmond}, {Rockosi}, {Schneider},
  {Stoughton}, {Tucker}, {Vanden Berk}, {Yanny}, {York}, \& {SDSS
  Collaboration}}]{Tegmark2004}
{Tegmark}, M., {Blanton}, M.~R., {Strauss}, M.~A., {et~al.} 2004, \apj, 606,
  702

\bibitem[{{Tolstoy} {et~al.}(2009){Tolstoy}, {Hill}, \& {Tosi}}]{Tolstoy2009}
{Tolstoy}, E., {Hill}, V., \& {Tosi}, M. 2009, Annual Review of Astronomy and
  Astrophysics, 47, 371

\bibitem[{{Vasiliev}(2018)}]{Vasiliev2018}
{Vasiliev}, E. 2018, \mnras, 2556

\bibitem[{{Vasiliev} \& {Valluri}(2020)}]{Vasiliev2020}
{Vasiliev}, E., \& {Valluri}, M. 2020, \apj, 889, 39

\bibitem[{{Walker} {et~al.}(2009{\natexlab{a}}){Walker}, {Mateo}, \&
  {Olszewski}}]{W09AJ}
{Walker}, M.~G., {Mateo}, M., \& {Olszewski}, E.~W. 2009{\natexlab{a}}, \aj,
  137, 3100

\bibitem[{{Walker} {et~al.}(2009{\natexlab{b}}){Walker}, {Mateo}, {Olszewski},
  {Pe{\~n}arrubia}, {Evans}, \& {Gilmore}}]{W09ApJ}
{Walker}, M.~G., {Mateo}, M., {Olszewski}, E.~W., {et~al.} 2009{\natexlab{b}},
  \apj, 704, 1274

\bibitem[{{Walker} {et~al.}(2010){Walker}, {Mateo}, {Olszewski},
  {Pe{\~n}arrubia}, {Evans}, \& {Gilmore}}]{Werr}
---. 2010, \apj, 710, 886

\bibitem[{{Wu}(2007)}]{Wu2007}
{Wu}, X. 2007, arXiv e-prints, astro

\bibitem[{{Zhao}(1996)}]{ZhaoDM}
{Zhao}, H. 1996, \mnras, 278, 488

\bibitem[{{Zwicky}(1933)}]{Zwicky1933}
{Zwicky}, F. 1933, Helvetica Physica Acta, 6, 110

\end{thebibliography}
\bibliographystyle{aasjournal}



\appendix
\section{More details about the {\bf modeling} results}
\subsection{NEMO results} \label{NEMO_detailed}

\begin{figure}
\caption{Models velocity dispersion $\chi^2$.
Horizontal lines are the mean values of $\chi^2$ for
times $t\geq 0.35$ for NEMO model~1 and for times $t\geq 0.2$
for AGAMA models~2 and 3.}
\label{chisq_tot}
\includegraphics[height=0.3\textheight]{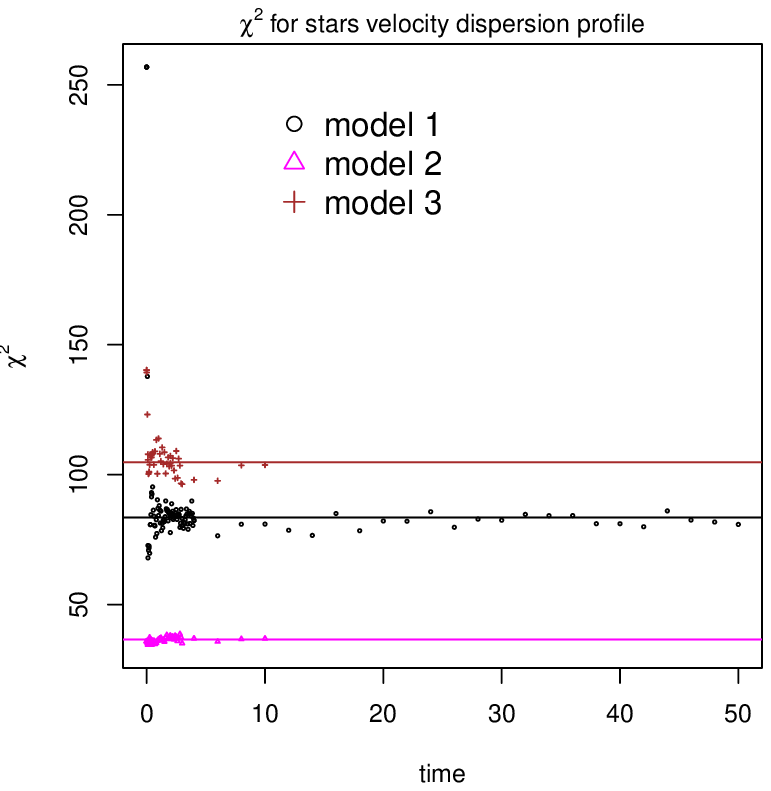}
\includegraphics[height=0.3\textheight]{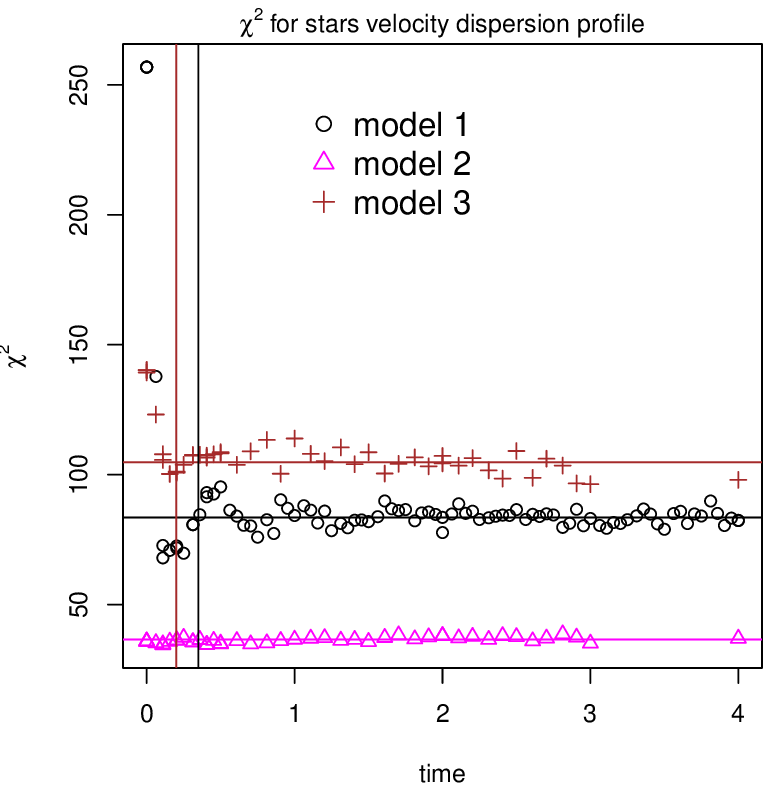}
\end{figure}

In Figure~\ref{chisq_tot}
variations of the $\chi^2$ values
over time are depicted.
We can see how the $\chi^2$ moves towards it's equilibrium value
for NEMO model~1 for times $t<4$
in the right panel of Figure~\ref{chisq_tot}.
This equilibrium value of $\chi^2$ can be estimated
as the mean values for times about $t\geq 0.35$
and is depicted as horizontal black line in Figure~\ref{chisq_tot}.
The estimations are included in Table~\ref{chisq_tot}.
We can see that the $\chi^2$ quantity starts to oscillate over its
equilibrium value much more earlier than the $\Delta$ quantity
(see {\bf Figure}~\ref{virial_ratios}),
so we choose this time interval for averaging of $\chi^2$ for the NEMO model.

\subsection{AGAMA results} \label{AGAMA_detailed}

\begin{figure}
\caption{Evolution of the virial ratio for AGAMA models}
\label{virial_ratio_AGAMA} 
\includegraphics[height=0.3\textheight]{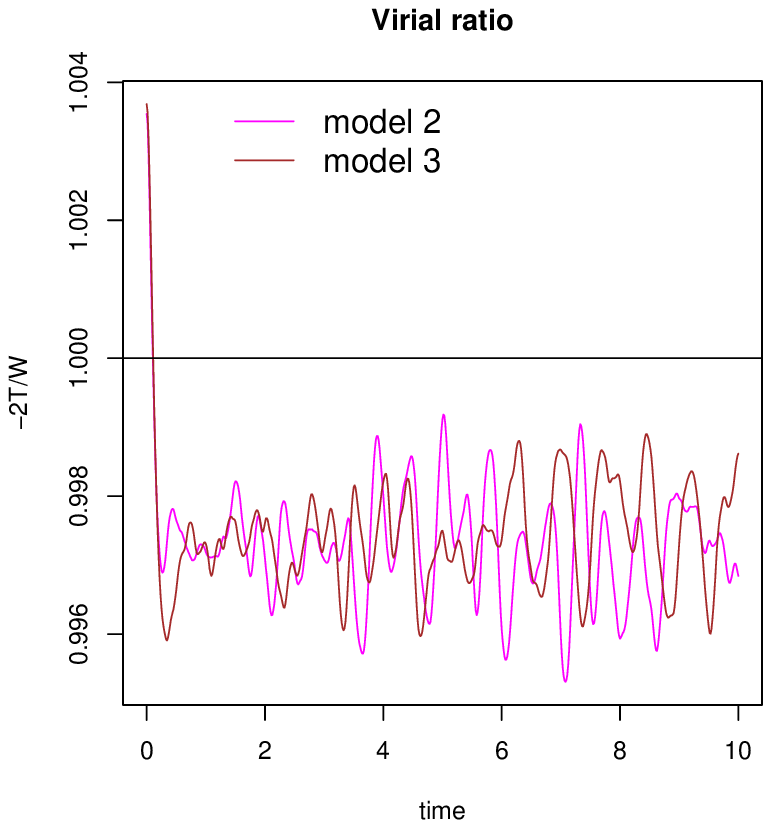}
\includegraphics[height=0.3\textheight]{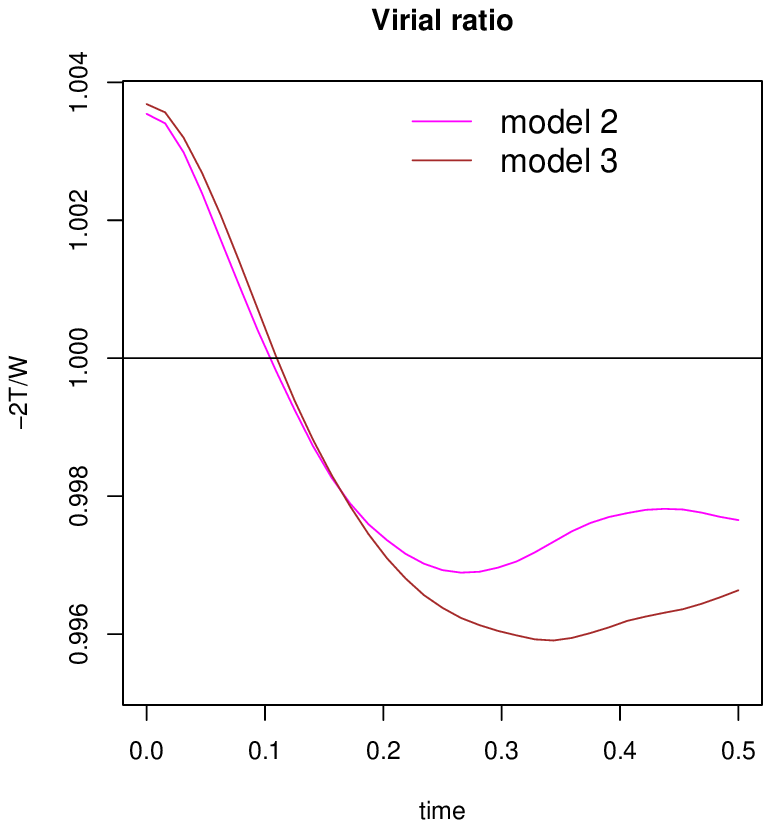}
\end{figure}

Let us consider in detail the equilibrium characteristics of our AGAMA models.
Their virial ratio evolution is depicted in Figure~\ref{virial_ratio_AGAMA}.
We can see that the absolute magnitude of $\Delta$ start values are not greater
than their following evolution values in contrast to our NEMO model.
But the sign of violation for all our N-body systems starts from positive one
tending to negative value during the first time interval of the evolution,
see Figure~\ref{virial_ratio_AGAMA}. In terms of equilibrium all our
AGAMA models are equally good.

Let us look at the evolution of the $\chi^2$ values for the models~2--3
in Figure~\ref{chisq_tot}
in order to compare their accordance with the data.
In the right panel of Figure~\ref{chisq_tot} we can see the initial
steps of the $\chi^2$ evolution. The setting of
equilibrium is completed by the time $t=0.2$
(see brown vertical line)
for model~3 and is not resolved for model~2.
So in Figure~\ref{chisq_tot}
the lines for the mean values of $\chi^2$ after $t=0.2$ 
for models~2 and 3 are depicted.
We find that the best model with the lowest $\chi^2$ is model~2
with the core DM density profile and the King best-fit 
stellar component parameters.

\section{King distribution}
In this appendix we give a summary of a usefull formulas for the~\citet{K62}
distribution which are not easily find in the textbook.

\subsection{Central surface density of the King's distribution}
\label{King_surface}
Now we shall express characteristic density $\rho_b$
in terms of observed quantities $r_c, t$ and $\Sigma_0$~-- central
surface density of the Fornax galaxy. Then we are to find the expression
of $\Sigma_0$ for the King's distribution:
\begin{eqnarray}
\Sigma_0 \equiv 2\int_0^{r_t} \rho_{\rm King}(r)dr \equiv
2 \rho_b r_c s_0(t) \; ,
\end{eqnarray}
where
\begin{eqnarray}
s_0(t) =
\arctan t - \frac{2 \log\left(t + \sqrt{t^2+1}\right)}{\sqrt{t^2+1}}
+ \frac{t}{t^2+1} \; .
\end{eqnarray}
For the characteristic density $\rho_b$ expressed in terms of $r_c, t, \Sigma_0$
we get:
\begin{eqnarray}
\label{rhob_inTermsof_Sigma}
\rho_b = \frac{\Sigma_0}{2 r_c s_0(t)} \; .
\end{eqnarray}

\section{Gravitational potential from axisymmetric density distribution (oblate systems)}
In this appendix we cite some formulae from~\citet{BT} in order to write
down equations for the radial distribution of the potential at the equatorial plane
for the Plummer oblate density profile and DM oblate density profiles. The latter
formulae are usefull for the radial distribution of the potential for the prolate DM density
profiles. We need these distributions for fitting NEMO code input parameters.

Let us consider gravitational system with equidensity axisymmetric ellipsoids of ellipticity $e$
and axial ratio $Q$. Then for our parameter we take $m$, so, that for cylindrical coordinates
of equidensity ellipsoid $(R,z)$ we have equation:
\begin{eqnarray}
\label{m_definition}
&m^2 = R^2 + z^2/Q^2, \; &Q^2 = 1 - e^2 .
\end{eqnarray}
Then we can define density as a function of $m$, $\rho(m)$, and define function $\mu(m)$ as:
\begin{eqnarray}
\label{Psi_definition}
\mu(m) = \int_0^m \rho(\bar m^2) d(\bar m^2).
\end{eqnarray}

Then from~\citet{BT} we have a formula for gravitational potential $\Psi(R_0,z_0)$ of such systems.
Let us write down the rise of the potential over its central value in terms of $Q$ instead of $e$:
\begin{multline}
\label{dPhi_Q}
\Psi(R_0,z_0) - \Psi(0,0) =
 \pi G Q a_0
\int_0^\infty \frac{\mu(m)d\tau}{(\tau+{a_0}^2)\sqrt{\tau+{b_0}^2}} \; ,\\
\mbox{where }\tau\mbox{ is defined by the equation:} \;
\frac{{R_0}^2}{\tau+{a_0}^2} + \frac{{z_0}^2}{\tau+{b_0}^2} = \frac{m^2}{{a_0}^2}, \; b_0 = Q a_0  .
\end{multline}

\subsection{Plummer potential distribution at the equatorial plane}
Let's find potential of the prolate Plummer distribution~(\ref{Plum_dist}):
\begin{equation*}
\begin{cases}
&\rho_{\rm Plummer}(R,z)=\rho_p(m_\star)= \frac{3 M_p}{4 \pi {b_p}^3} \left[ 1 + \frac{{m_\star}^2}{{b_p}^2} \right]^{-5/2} \text{, } \\
&{m_\star}^2 = R^2 + \frac{z^2}{q^2}\text{ ,}
\end{cases}
\end{equation*}
with the oblateness $q$ and ellipticity $e=\sqrt{1-q^2}$.
Then the function $\mu$ from~(\ref{Psi_definition}) for this profile will be:
\begin{eqnarray}
\mu(m) = \frac{M_p}{2\pi b_p} \left(
                    1 - \frac{1}{\left( 1 + m^2/{b_p}^2\right)^{3/2}}
                               \right) \; .
\end{eqnarray}
Let's find the potential distribution at the equatorial plane
for $z_0=0$:
\begin{eqnarray}
\label{Plummer_phi}
\Psi(R_0,0.0) = - \frac{M_p G q}{b_p \sqrt{1-q^2}}
\times \left.\left( \frac{x}{\left( k + 1 \right)^{3/2}}
               F_1 \left( \frac{1}{2};-\frac{1}{2};\frac{3}{2};\frac{3}{2};-x^2;-\frac{x^2}{k+1}\right)
       \right)\right|_{q/e}^{\infty} ,
\end{eqnarray}
where $F_1$ is Gauss hypergeometric function.

\subsection{Plummer mass for the oblate system}
In order to calculate the mass of such a system
we shall use formulae for mass of a thin homeoid
between ellipsoids
with oblateness $q$ and equatorial radii $m$ and $m+dm$
from~\citet{BT}:
\begin{equation}
\delta M = 4 \pi \rho(m_\star) {m_\star}^2 q dm \; .
\end{equation}
Then the mass inside such an ellipsoid will be equal to:
\begin{eqnarray}
\label{Plum_oblate_mass}
M(m_\star) =
  4 \pi q \int_0^{m_\star} \rho(\tilde{m}) \tilde{m}^2 d \tilde{m}
           = q M_p \frac{m^3}{({b_p}^2 + m^2)^{3/2}} \; .
\end{eqnarray}

\subsection{Zhao density distribution with one parameter}
\label{density_Kohey}
Let us consider~\citet{H16}
density distribution of the DM halo.
More common form of such a DM profile we can find at~\citet{ZhaoDM}.
We shall rewrite the
function $\rho(m)$ from equation~(\ref{rhoDM_distribution}) in terms of variable $p=m/b_{\rm halo}$:
\begin{eqnarray}
\label{p_var}
\rho(p) = \rho_0 p^\alpha (1+p^2)^{-(\alpha+3)/2} \; , \;
p=\frac{m}{b_{\rm halo}}, \; \tilde p=\frac{\bar m}{b_{\rm halo}};
\end{eqnarray}
and calculate function $\mu(m)$ for our density distribution:
\begin{eqnarray}
\label{Psi_dep_p}
\mu(m) =
\frac{\rho_0 b_{\rm halo}^2}{\frac{\alpha}{2}+1} \tilde{p}^{\alpha+2}
2F_1(\frac{\alpha}{2}+1, \frac{\alpha}{2}+1.5, \frac{\alpha}{2}+2, -\tilde{p}^2) \bigg|_0^\frac{m}{b_{\rm halo}} ,
\end{eqnarray}
where $2F_1(a,b,c,z)$ is Gauss hypergeometric function.

\section{Potential of a thin prolate ellipsoidal shell}
Let us derive analogue of formulas~(\ref{dPhi_Q}) for prolate systems,
e.i. equidensity axisymmetric ellipsoids with axial ratio $Q>1$.
First we shall examine a thin prolate ellipsoidal shell
and than summarise the contribution of such shells into the total potential.
By the definition the axial ratio of the shell is $Q$
and it has a coordinate $m$, denoting the shell coordinates $R_\star, z_\star$
as in equation~(\ref{m_definition}):
\begin{eqnarray}
\label{m_prolate}
{R_\star}^2 + \frac{{z_\star}^2}{Q^2} = {m}^2
\end{eqnarray}
Let us define the prolate coordinates as in problem 2.5 of Chapter 2~\citet{BT}
\begin{eqnarray}
\label{prolate_def}
R = a \sinh u \sin v, \; z = a \cosh u \cos v \; , \;
\phi \mbox{ -- azimuthal angle} \; .
\end{eqnarray}
For 	comparison the oblate coordinates for $Q<1$ are:
$R =a \cosh u \sin v, z = a \sinh  u \cos v$, $\phi$~-- the same azimuthal angle.
For our coordinates we can express $\sin v$ and $\cos v$ from~(\ref{prolate_def})
and use the trigonometrical identical equation:
\begin{eqnarray}
\label{Rz_uv}
\frac{R^2}{\sinh^2 u} + \frac{z^2}{\cosh^2 u} = a^2 \; .
\end{eqnarray}

We are defining our prolate coordinates so, that one of the ellipsoids
with $u=const$ coincide
with our thin shell that generates gravitation potential $\delta\Psi$.
Let us denote coordinate for this shell as $u_\star$.
Than comparing equations~(\ref{prolate_def}) and~(\ref{Rz_uv}) for
$R_\star, z_\star, u_\star$ we can write for $a$ and $u_\star$:
\begin{eqnarray}
\label{a_u_star}
\coth u_\star = Q, \;
a = m \csc u_\star = m \sqrt{Q^2 - 1} \; .
\end{eqnarray}
All our ellipsoids with $u=const$ are confocal with the focus $a=m \sqrt{Q^2 - 1}$. We can see that our coordinates are suitable for $Q>1$, because
$\coth u_\star > 1$ for all positive $u_\star$. And for oblate coordinates
it will be $\tanh u_\star = Q < 1$ for all positive $u_\star$.

In order to find the potential $\delta\Psi$ we are to solve the equation $\nabla^2 \delta \Psi = 0$.
We shall find the expression for $\nabla^2 \delta \Psi$ in our prolate coordinates.
Now we are varying in turn coordinates $u, v, \phi$ and moving along orthogonal
vectors ${\bf \hat e}_u, {\bf \hat e}_v, {\bf \hat e}_\phi$ by the distances
$h_u \delta u$, $h_v \delta v$,$h_\phi \delta \phi$:
\begin{eqnarray}
\label{h_prolate}
h_u=
     a \sqrt{
              \cosh^2 u \sin^2 v + \sinh^2 u \cos^2 v
             } \; ,\;
h_v=
     a \sqrt{
              \cosh^2 u \sin^2 v + \sinh^2 u \cos^2 v
             } \; , \;
h_\phi= R =  a \sinh u \sin v \; .
\end{eqnarray}
After some transformations we have for $h_u, h_v$:
\begin{eqnarray}
h_u = h_v = a \sqrt{\cosh^2 u - \cos^2 v} \; .
\end{eqnarray}
For 	comparison we can write down the coefficients for oblate systems
in oblate coordinates:
$h_u=h_v= a \sqrt{\sinh^2 u + \cos^2 v}$, $h_\phi = a \cosh u \sin v$.
Following the equation for gradient:
\begin{eqnarray}
\label{grad}
\nabla = \frac{{\bf \hat e}_i}{h_i} \frac{\partial}{\partial q_i}, \;
q_1 = u, \; q_2 = v, \; q_3 = \phi .
\end{eqnarray}
we shall write down the gradient of potential $\delta\Psi$:
\begin{eqnarray}
\label{grad_dPhi}
\nabla \left( \delta \Psi \right) = \frac{1}{a\sqrt{\cosh^2 u - \cos^2 v}}
\left(
       \frac{\partial \left(\delta\Psi\right)}{\partial u} {\bf \hat e}_u +
       \frac{\partial \left(\delta\Psi\right)}{\partial v} {\bf \hat e}_v
\right)
+ \frac{1}{a \sinh u \sin v} \frac{\partial\left( \delta\Psi\right)}{\partial \phi}
{\bf \hat e}_\phi \; .
\end{eqnarray}
The equation for Laplacian is:
\begin{eqnarray}
\label{Laplace}
\nabla^2 F = \frac{1}{h_1 h_2 h_3}
\left(
      \frac{\partial}{\partial q_1} \left(
             \frac{h_2 h_3}{h_1} \frac{\partial F}{\partial q_1}
                                    \right) +
      \frac{\partial}{\partial q_2} \left(
             \frac{h_3 h_1}{h_2} \frac{\partial F}{\partial q_2}
                                    \right)
+ \frac{\partial}{\partial q_3} \left(
             \frac{h_1 h_2}{h_3} \frac{\partial F}{\partial q_3}
                                    \right)
\right) \; .
\end{eqnarray}
For $h_1=h_2$ and $h_3$ all independent of $q_3$ we can rewrite this
equation in the following way:
\begin{eqnarray}
\label{Laplace_q3indipend}
\nabla^2 F = \frac{1}{{h_1}^2 h_3} \left(
                                  \frac{\partial}{\partial q_1} \left(
                                       h_3 \frac{\partial F}{\partial q_1}
                                                                \right) +
                                  \frac{\partial}{\partial q_2} \left(
                                       h_3 \frac{\partial F}{\partial q_2}
                                                                \right)
                                      \right)
+ \frac{1}{{h_3}^2} \frac{\partial^2 F}{\partial^2 q_3} \; .
\end{eqnarray}
Then for the Laplacian of function $\delta\Psi$ we can write down:
\begin{multline}
\label{Laplace_dPhi}
\nabla^2 \left(\delta \Psi \right)
=  \frac{1}{a^2(\sinh^2 u - \cos^2 v)} \left(
  \frac{1}{\sinh u} \frac{\partial}{\partial u} \left(
                                 \sinh u \frac{\partial \left( \delta \Psi\right)}{\partial u}
                                                \right)
+ \frac{1}{\sin v} \frac{\partial}{\partial v} \left(
                                  \sin v \frac{\partial \left( \delta \Psi\right)}{\partial v}
                                               \right)
                                        \right) \\
 +\frac{1}{a^2 \sinh^2 u \sin^2 v}
  \frac{\partial^2 \left( \delta \Psi\right)}{\partial^2 \phi} \; .
\end{multline}

We are interested in the potentials, constant over the $\phi$ and $v$
variables, so we can write down:
\begin{eqnarray}
\label{dPhi_pShellAB}
\begin{cases}
 \frac{\partial \left( \delta \Psi\right)}{\partial v} = 0 , \\
 \frac{\partial \left( \delta \Psi\right)}{\partial \phi} =0 ,\\
 \nabla^2 \left( \delta \Psi\right) = 0 .
\end{cases}
\Rightarrow
\frac{d}{du} \left(
                  \sinh u \frac{d \left(\delta \Psi\right)}{du}
             \right) = 0
\Leftrightarrow
\frac{d \left( \delta \Psi \right)}{du} = \frac{A}{\sinh u} \; , \;
\delta \Psi = A \log \left( \tanh \frac{u}{2} \right) + B \; .
\end{eqnarray}

In order to find constants $A$ and $B$ we are to take the limit
of function $\delta\Psi$ at the infinite value of variables $R$ and $z$,
e.i. infinite $r=\sqrt{R^2 + z^2}$.
The limit of the potential of our ellipsoidal shell should be:
\begin{eqnarray}
\label{dPhi_lim}
\lim_{r\to\infty} \delta \Psi = - \frac{G \delta M}{r} \; ,
\end{eqnarray}
where $\delta M$ is the mass of the ellipsoidal shell.

Let us express $r$ in terms of $u$ and $v$ variables:
\begin{eqnarray}
\label{r_of_uv}
r^2 \equiv R^2 + z^2 =a^2 (\sinh^2 u + \cos^2 v) \; ,
\end{eqnarray}
and find $\sinh u$ from this expression:
\begin{eqnarray}
\label{sh_u_p}
a^2 \sinh^2 u = r^2 - a^2 \cos^2 v \; .
\end{eqnarray}
Now subtracting~(\ref{sh_u_p}) from $a^2 \cosh^2 u$
we can use the hypergeometric identical equation:
\begin{eqnarray}
\label{ch_u_p}
a^2 \cosh^2 u = r^2 + a^2 \sin^2 v \; .
\end{eqnarray}
Then we shall rewrite $\tanh \dfrac{u}{2}$ in terms of $r$ and $v$,
using equation~(\ref{ch_u_p}):
\begin{eqnarray}
\log \left(
            \tanh \frac{u}{2}
     \right)
= \frac{1}{2} \log \left(
  \frac{\sqrt{r^2 + a^2 \sin v} - a}{\sqrt{r^2 + a^2 \sin v} + a}
                    \right) \; .
\end{eqnarray}
Now we shall take a Taylor  series expansion of $\delta \Psi$
for small $\dfrac{a}{r}$:
\begin{eqnarray}
\label{dPhi_pShall_Taylor}
\delta\Psi = - A \frac{a}{r} + B + o\left( \frac{a}{r} \right) \; ,
\end{eqnarray}
To get the limit of $\delta \Psi$ same as in equation~(\ref{dPhi_lim})
we can write down for constants $A$ and $B$:
\begin{eqnarray}
A =  \frac{G \delta M}{a} \; , B = 0 \; .
\end{eqnarray}

For the points inside our shell the potential is constant and
equal to its value on the shell itself. Let us do some
hypertrigonometric transformations to get this value:
\begin{eqnarray}
\log \left(
           \tanh \frac{u_\star}{2}
     \right)
= \log \left(
               Q - \sqrt{Q^2 - 1}
         \right) \; .
\end{eqnarray}

So, the final result for the potential of thin prolate ellipsoidal shell is:
\begin{eqnarray}
\label{dPhi_pShell_fin}
\delta \Psi =
\begin{cases}
\frac{G\delta M}{m\sqrt{Q^2-1}}\log\left(\tanh\frac{u}{2}\right), &\text{ $u>u_\star$, outside;}\\
\frac{G\delta M}{m\sqrt{Q^2-1}}\log\left(Q-\sqrt{Q^2-1}\right),&\text{$u \le u_\star$, inside.}
\end{cases}
\end{eqnarray}

Let us find the mass $\delta M$ of our shell with the density $\rho(m)$.
The semi-axis of the ellipsoid are $m$ and $Qm$, so the total volume
inside the ellipsoid is:
\begin{eqnarray}
V = \frac{4}{3} \pi m^3 Q \; ,
\end{eqnarray}
and the volume $\delta V$ inside the shell is:
\begin{eqnarray}
\label{deltaVp}
\delta V
2 \pi Q m \delta \left( m^2 \right).
\end{eqnarray}
Then the mass $\delta m$ of the shell is:
\begin{eqnarray}
\label{deltaMp}
\delta M = 2 \pi Q \rho(m) m \delta \left( m^2 \right).
\end{eqnarray}
Now we can rewrite the equation~(\ref{dPhi_pShell_fin})
for the potential of thin prolate ellipsoidal shell
using the expression for $\delta M$:
\begin{eqnarray}
\label{dPhi_pShell_m}
\delta \Psi =
\begin{cases}
2\pi G \rho(m)\frac{Q}{\sqrt{Q^2-1}}
\log\left(\tanh\frac{u}{2}\right)\delta \left( m^2 \right),
&\text{ $u>u_\star$;}\\
2\pi G \rho(m)\frac{Q}{\sqrt{Q^2-1}}
\log\left(Q-\sqrt{Q^2-1}\right)\delta \left( m^2 \right),
&\text{$u \le u_\star$;}
\end{cases}
\end{eqnarray}
where $u_\star$ is the coordinate of the shell
and $u$ is the coordinate of the ellipsoid, that is confocal with the
thin shell and passes through the point ($R_0,z_0$).

\subsection{Gravitational potential for prolate systems}
Now we are ready to summarise the potentials of prolate ellipsoidal shells
at the point $(R_0, z_0)$ to get the total potential $\Psi$.
For each shell we have coordinates $m$, $R_\star, z_\star$, $u_\star$
of the shell itself, which satisfy the equation~(\ref{Rz_uv}),
and coordinate of the ellipsoid $u_m$ on which the point $R_0, z_0$ lies,
and this ellipsoid is confocal with the shell. So, the coordinates
$u_m, R_0, z_0$ also satisfy the equation~(\ref{Rz_uv}):
\begin{eqnarray}
\label{R0_Z0_um}
\frac{{R_0}^2}{\sinh^2 u_m} + \frac{{z_0}^2}{\cosh^2 u_m} = m^2 \left( Q^2 - 1 \right)
\end{eqnarray}

Our point $(R_0, z_0)$ lies on the equidensity shell with the coordinate
$m_0$. Then the point $(R_0, z_0)$ lies outside the shell which
coordinate $m < m_0$ and inside the shell which coordinate $m > m_0$.
Let us denote as $\delta \Psi_{out}(m)$ the contribution
to the total potential $\Psi$ of that shells,
for which the point $(R_0, z_0)$ lies outside the shell.
Then for such shells
our coordinate $u_m$ is grater then the shell coordinate $u_\star$:
$u_m > u_\star$.
The  contribution of shells,
for which the point $(R_0, z_0)$ lies inside the shell will be denoted as
$\delta \Psi_{in}(m)$.  Then for this shells
our coordinate $u_m$ is less then the shell coordinate $u_\star$:
$u_m < u_\star$.
And the sum of $\delta \Psi_{in}(m)$ is the case
$u \le u_\star$ in the formula~(\ref{dPhi_pShell_m}),
and the sum of $\delta \Psi_{out}(m)$  is the case
$u > u_\star$ in that formula.
We shall write down the sum of $\delta \Psi_{in}(m)$ taking into account
the definition of function $\mu(m)$~(\ref{Psi_definition}):
\begin{eqnarray}
\label{deltaPhi_in_p}
\sum_{m>m_0} \delta \Psi_{in}(m) =
 2 \pi G \frac{Q}{\sqrt{Q^2-1}} \left(
                                      \mu(\infty) - \mu(m_0)
                                 \right)
                                  \log \left(
                                          Q - \sqrt{Q^2 - 1 }
                                      \right) .
\end{eqnarray}
The sum of $\delta \Psi_{out}(m)$ will be:
\begin{eqnarray}
\label{deltaPhi_out_p}
\sum_{m<m_0}\delta \Psi_{out}(m)
=  2 \pi G \frac{Q}{\sqrt{Q^2-1}}
  \left(
 \mu(m) \log \left(
                \tanh \frac{u_m}{2}
              \right)\bigg|_0^{m_0} \right.
- \left. \int_0^{m_0} \mu(m) d \left(
                        \log \left( \tanh \frac{u_m}{2} \right)
                        \right)
  \right)
\end{eqnarray}
Let us rewrite the differential:
\begin{eqnarray}
\label{dlog_th}
d \left(
      \log \left( \tanh \frac{u_m}{2} \right)
  \right)=
\frac{d \cosh u_m}{\sinh^2 u_m}
\end{eqnarray}

For $m=m_0$ ellipsoid $u_m$ coincides with the ellipsoid $u_\star$
and we can write $\coth u_m=Q$ as in the equation~(\ref{a_u_star}).
So we can evalute the  $\cosh u_m = \frac{Q}{\sqrt{Q^2-1}}$ for $m=m_0$.
And for $m=0$ to satisfy the equation~(\ref{Rz_uv}) we are to set $\coth u_m = \infty$.
Summarising equations~(\ref{deltaPhi_in_p}) and~(\ref{deltaPhi_out_p})
and taking into account the differential~(\ref{dlog_th})
we can derive the gravitational potential at the point $\Psi(R_0,z_0)$:
\begin{eqnarray}
\label{Phi_prolate}
\Psi(R_0,z_0) =  2 \pi G \frac{Q}{\sqrt{Q^2-1}}
\left( \mu(\infty) \log\left( Q - \sqrt{Q^2-1} \right) \right.
  + \left. \int_{\cosh u_m = \frac{Q}{\sqrt{Q^2-1}}}^{\infty} \mu(m) \frac{d \cosh u_m}{\sinh^2 u_m} \right) \; .
\end{eqnarray}
Let us rewrite the expressions for $\sinh u_m, \cosh u_m$ in terms
of $\tau, a_0, b_0$:
\begin{multline}
\label{a0_b0_tau_prolate}
                   b_0 = Q a_0 \; ; \;
        \tau + {a_0}^2 = {a_0}^2 \sinh^2 u_m \left( Q^2 - 1 \right) \;
                       \Rightarrow \;
{a_0}^2 \cosh^2 u_m
\left( Q^2 - 1 \right) =
\tau + {b_0}^2 \; , \\
d \tau   = 2 a_0 \sqrt{Q^2-1} \sqrt{\tau+{b_0}^2}d \cosh u_m \; .
\end{multline}
For $m=m_0$ we can write $\sinh u_m = \frac{1}{\sqrt{Q^2-1}}$
and than $\tau=0$. For $m=0$ and $\cosh u_m = \infty$
we can write $\tau=\infty$.
Then the rise of the potential over its central value could be write down as:
\begin{eqnarray}
\label{dPhi_prolate}
\Psi(R_0,z_0) - \Psi(0,0)  =
 \pi G Q a_0 \int_0^{\infty} \mu(m)
\frac{d \tau}{(\tau+{a_0}^2)\sqrt{\tau + {b_0}^2}} \; .
\end{eqnarray}
We can see that the definitions of $a_0, b_0$ and the equations~(\ref{dPhi_Q}) and~(\ref{dPhi_prolate}) for
$\Psi(R_0,z_0) - \Psi(0,0)$ are identical for oblate and prolate systems!

\section{Notations for program codes parameters}
\label{code_notations}
\begin{deluxetable}{cccccccc}
\tablecaption{Code parameters for NEMO mkkd95 runs.\label{NEMO_code} }
\tablehead{
\colhead{\bf q} & \colhead{\bf psi0} & \colhead{\bf v0} & \colhead{\bf ra} & \colhead{\bf rck2} & \colhead{\bf rhob} & \colhead{\bf psicut } & \colhead{\bf sigb }
}
\startdata
$q$ & $\Psi_0$ & $v_0$ & $R_a$ & ${r_{ck}}^2$ & $\rho_b$ & $\Psi_c$ & $\sigma_b$ \\
\enddata
\end{deluxetable}

We can see the vocabulary for program codes parameters
in the following tables~\ref{NEMO_code} and~\ref{AGAMA_code}.
The fist row of the tables are the codes parameters in bold font
and the second row are the Greek letters notations for the
parameters used in the paper.
Table~\ref{NEMO_code} shows the NEMO mkkd95 code parameters.
Table~\ref{AGAMA_code} shows the AGAMA Schwarzschild's {\bf modeling} code parameters.
The notation for falcON code parameters are the follows:
\begin{itemize}
\item {\bf kmax}~-- $k_{max}$,
\item  {\bf eps}~-- $\epsilon$.
\end{itemize}

\begin{splitdeluxetable}{CCCCCCCCBCCCCCCC}
\tablecaption{Code parameters for AGAMA Schwarzschild's runs.\label{AGAMA_code}}
\tablehead{
\multicolumn{8}{c}{DM parameters}&\multicolumn{7}{c}{Stellar parameters}\\
\colhead{\bf densityNorm}&\colhead{\bf mass}&\colhead{\bf scaleRadius}&\colhead{\bf alpha}&\colhead{\bf beta} &\colhead{\bf gamma}&\colhead{\bf axisRatioY}&\colhead{\bf axisRatioZ}&\colhead{\bf W0}&\colhead{\bf mass}&\colhead{\bf scaleRadius}&\colhead{\bf beta}&\colhead{\bf icbeta} &\colhead{\bf ickappa}&\colhead{\bf axisRatioZ}
}
\startdata
             \rho_0      &    M_{\rm DM}    &         b_{\rm halo}    &        \gamma     &         \eta      &         -\alpha   &         p              &           Q            &       W_0      & M_{\rm stars}    &        b_p, r_{cA}         &        \beta     &        \beta_z      &         \kappa      &        q               \\
\enddata
\end{splitdeluxetable}

\end{document}